%% file: 19TWC-LinOFDM_arxiv.tex
\documentclass[journal,twoside,twocolumn]{IEEEtran}
\usepackage{cite}
\usepackage[T1]{fontenc}
\usepackage{graphicx}
\usepackage{amssymb}
\usepackage{amsmath}
\usepackage{paralist}
\usepackage{setspace}
\usepackage{microtype}
\usepackage{hyperref}
\usepackage{url}
\usepackage{balance}
\usepackage[caption=false,font=footnotesize]{subfig}
\usepackage{xcolor}
\usepackage{fancyref}
\usepackage{dsfont}
\usepackage{setspace}

\newcommand{\quantize}{\mathcal{Q}}

\newcommand{\rounding}{{\mathcal{R}}}

\newcommand{\snr}{\rho}

\newcommand{\sindr}{\gamma}
\newcommand{\osr}{\xi}

\input{vmr-symbols-vecbold}
\input{standard-macros}

\input{defs}

\graphicspath{{./}}
\newcommand{\figwidth}{.475\textwidth}  

\begin{document}

%
\title{Linear Precoding with Low-Resolution DACs \\ for Massive MU-MIMO-OFDM Downlink}%
\author{%
Sven Jacobsson~\IEEEmembership{Student~Member,~IEEE}, Giuseppe Durisi~\IEEEmembership{Senior~Member,~IEEE}, Mikael Coldrey, and \\Christoph Studer~\IEEEmembership{Senior~Member,~IEEE}%
\thanks{S.\ Jacobsson is with Ericsson Research, 41756 Gothenburg, Sweden and~also with the Department of Electrical Engineering, Chalmers University of Technology, 41296 Gothenburg, Sweden (e-mail: \url{sven.jacobsson@ericsson.com}).} 
\thanks{G.\ Durisi is with the Department of Electrical Engineering, Chalmers University of Technology, 41296 Gothenburg, Sweden (e-mail: \url{durisi@chalmers.se}).}
\thanks{M.\ Coldrey is with Ericsson Research, 41756 Gothenburg, Sweden (e-mail: \url{mikael.coldrey@ericsson.com}).} 
\thanks{C.~Studer is with the School of Electrical and Computer Engineering, Cornell University, Ithaca, NY 14853, USA (e-mail: \url{studer@cornell.edu}; web: \url{http://vip.ece.cornell.edu}).}
\thanks{The work of S.~Jacobsson and G.~Durisi was supported by the Swedish Foundation for Strategic Research under grant ID14-0022, and by the Swedish Governmental Agency for Innovation Systems (VINNOVA) within the competence center ChaseOn. The work of C.~Studer was supported in part by Xilinx, Inc.\ and by the US National Science Foundation (NSF) under grants ECCS-1408006, CCF-1535897, CAREER CCF-1652065, and CNS-1717559.}
\thanks{The authors would like to thank  D.~Astely and U.~Gustavsson at Ericsson Research for fruitful discussions.}
\thanks{The MATLAB simulator used to obtain the numerical results in~\fref{sec:numerical} is available on GitHub: \url{https://github.com/quantizedmassivemimo/1bit_linear_precoding_ofdm}.}
\thanks{This paper was presented in part at the IEEE Global Commun.~Conf.~(GLOBECOM), Singapore, Dec.~2017~\cite{jacobsson17c}.}
}
\maketitle

\vspace{-.825cm}


\begin{abstract}

We consider the downlink of a massive multiuser (MU) multiple-input multiple-output (MIMO) system in which the base station (BS) is equipped with low-resolution digital-to-analog converters~(DACs). In contrast to most existing results, we assume that the system operates over a frequency-selective wideband channel and uses orthogonal frequency division multiplexing~(OFDM) to simplify equalization at the user equipments (UEs). Furthermore, we consider the practically relevant case of oversampling DACs.
We theoretically analyze the uncoded bit error rate (BER) performance with linear precoders (e.g.,~zero forcing) and quadrature phase-shift keying using Bussgang's theorem.
We also develop a lower bound on the information-theoretic sum-rate throughput achievable with Gaussian inputs, which can be evaluated in closed form for the case of 1-bit DACs. For the case of multi-bit DACs, we derive approximate, yet accurate, expressions for the distortion caused by low-precision DACs, which can be used to establish lower bounds on the corresponding sum-rate throughput. 
Our results demonstrate that, for a massive MU-MIMO-OFDM system with a 128-antenna BS serving 16 UEs, only 3--4~DAC bits are required to achieve an uncoded BER of $\boldsymbol{10^{-4}}$ with a negligible performance loss compared to the infinite-resolution case at the cost of additional out-of-band emissions. Furthermore, our results highlight the importance of taking into account the inherent spatial and temporal correlations caused by low-precision DACs.
\end{abstract} 


\section{Introduction}

Massive multiuser (MU) multiple-input multiple-output (MIMO) will be a key technology in future cellular communication systems~\cite{boccardi14a}. 
Massive MU-MIMO equips the base station (BS) with hundreds of active antenna elements to serve simultaneously tens of user equipments (UEs), which results in improved spectral efficiency and energy efficiency compared to traditional, small-scale MIMO systems~\cite{marzetta10a, rusek14a, larsson14a, lu14a}. Furthermore, all of these advantages can be achieved by means of simple~signal processing schemes (e.g., linear precoding) at the BS~\cite{marzetta10a, rusek14a}.

Increasing  the number of active antenna elements at the BS by orders of magnitude increases the circuit power consumption.
Data converters, i.e., analog-to-digital converters~(ADCs) and digital-to-analog converters~(DACs), are foreseen to be among the most dominant sources of power consumption in massive MU-MIMO systems. 
In today's state-of-the-art direct-conversion MIMO systems, a pair of high-resolution DACs (e.g., with $10$ bits of resolution per real dimension or more) is used to generate the {in-phase and quadrature components of the} transmitted baseband signal at each antenna element at the BS. Scaling such high-resolution architectures to massive MU-MIMO systems, in which the number of antennas at the BS could be in the order of hundreds, would lead to excessively high power consumption. 
Hence, to maintain a reasonable power budget, the resolution of the DACs must be reduced.
Furthermore, equipping the BS with hundreds of active antenna elements puts extreme data rate requirements on the interface connecting the baseband-processing unit to the radio unit (where the DACs are located), especially for wideband systems operating in the millimeter wave part of the wireless spectrum. By lowering the resolution of the DACs, one can---to some extent---mitigate this data rate bottleneck.
In this work, we consider the massive MU-MIMO \emph{downlink} (BS transmits data to multiple UEs) and focus on the case in which low-resolution DACs are used at the~BS.

\subsection{Relevant Prior Art}


\subsubsection{Transmitter Impairments and Out-of-Band Emissions} The impact of \emph{aggregate} radio frequency~(RF) hardware impairments (i.e., from multiple sources of hardware impairments) in the massive MU-MIMO downlink has been investigated in, e.g.,~\cite{gustavsson14a, bjornson14a, zhang15d}, using a model that treats hardware impairments as power-dependent additive white noise that is uncorrelated with the input signal.
Such models have been validated experimentally~\cite{studer10b}  under the assumptions that (i) 
the system uses orthogonal frequency-division multiplexing (OFDM)
and (ii) methods have been implemented at the BS to mitigate common hardware impairments.  
The latter indicates that the impairments in~\cite{gustavsson14a, bjornson14a, zhang15d} capture only \emph{residual} distortions that have not been entirely compensated.
Furthermore, such aggregate models are not useful for analyzing out-of-band~(OOB) emissions~\cite{larsson18a}, which are of significant concern in practice.
In general, distortions (in-band and OOB) caused by transmitter impairments are beamformed, to some extent, in the direction of the useful signal (i.e., they are correlated over the BS antenna array).
This behavior has been illustrated in, e.g.,~\cite{moghadam12a, mollen17a, blandino17a}, which analyzed the specific distortion caused by nonlinear power amplifiers~(PAs).
In~\cite{bjornson18a}, focusing again on the distortion caused by nonlinear PAs, it was shown that, in some cases, ignoring the spatial correlation leads to marginal errors when characterizing spectral efficiency.
As we shall see in~\fref{sec:numerical}, similar conclusions hold  for the distortion caused by DACs with medium-to-high resolution. 
However, for low-resolution (e.g., 1 bit) DACs,  the correlation of the distortion should not be ignored when characterizing bit-error rate (BER) and spectral~efficiency.

\subsubsection{Low-Resolution ADCs} The impact on performance of using low-resolution ADCs to quantize the received signal at the BS in the massive MU-MIMO \emph{uplink} (multiple UEs transmit to the BS) has been analyzed in, e.g., \cite{wen15b,studer16a, li17b, jacobsson17b, mollen16c}. 
These analyses suggest that reliable communication is possible even in the extreme case of 1-bit ADCs. Furthermore, low-resolution ADCs can be deployed at the BS with negligible performance degradation compared to the infinite-resolution (i.e., no quantization) case, provided that the ratio between the number of BS antennas and UEs  is sufficiently~large.
To facilitate an analytical performance analysis, simple approximate models that treat the distortion caused by the ADCs as a white additive noise that is uncorrelated with the input signal, have been used extensively in the literature (see, e.g.,~\cite{fan15a, orhan15a, sarajalic17a}).
Although these models are accurate under certain assumptions on, e.g., the resolution and the sampling rate of the converters, and on the number of active users, they fail to take into account the inherent spatial and temporal correlation of the quantization distortion~\cite{mezghani12b, li17e}. 
To assess the accuracy of these simplified models, we will develop a distortion model that takes this correlation into account and compare it to simplistic distortion~models.

\subsubsection{Low-Resolution DACs}
Linear precoders, such as maximal-ratio transmission (MRT), zero-forcing (ZF), and Wiener-filter~(WF) precoding for narrowband systems with $1$-bit DACs at the BS have been analyzed in, e.g., \cite{saxena16b, li17a, mezghani09c, jacobsson17d}. The results therein show that 1-bit-DAC massive MU-MIMO systems support low BERs and high sum-rate throughputs despite the severe nonlinearity caused by $1$-bit DACs.
Multi-bit DACs and linear precoding for the massive MU-MIMO downlink were considered in \cite{jacobsson17d}. There, it was shown that DACs with few bits (e.g., 3--4 bits) yield a performance that is close to the infinite-resolution case if the BS has access to perfect channel state information~(CSI). 
A linear WF-quantized~(WFQ) precoder was proposed in \cite{mezghani09c} on the basis of an approximate model for the distortion caused by the DACs. This precoder is shown to outperform conventional precoders for small-to-moderate MIMO systems at high signal-to-noise ratio~(SNR). However, as shown in \cite{jacobsson17d}, the performance gain of the WFQ precoder over conventional linear precoders~(e.g.,~ZF) is marginal in the massive MU-MIMO~case.

For the case of 1-bit DACs, more sophisticated nonlinear precoding strategies that significantly outperform linear precoders at the cost of an increased signal-processing complexity were presented in, e.g., \cite{jacobsson17d, jacobsson16d, jedda16a, swindlehurst17a, castaneda17a, landau17a, jacobsson18b, li18a, shao18a, shao18b, sohrabi18a}. 

All existing results on low-resolution DACs reviewed so far dealt with the frequency-flat case. The first work to consider 1-bit DACs combined with~(OFDM) for massive MU-MIMO  over frequency-selective channels is \cite{guerreiro16a}. 
There, by using an approximate model that treats the 1-bit-DAC distortion as white, it is shown that simple MRT precoding results in high signal-to-interference-noise-and-distortion ratio~(SINDR) at the UEs provided that the number of BS antennas is significantly larger than the number of UEs.
In the conference version of this paper~\cite{jacobsson17c}, we used an exact model for the 1-bit-DAC distortion to characterize the BER achievable in an OFDM system with quadrature phase-shift keying~(QPSK), as well as the sum-rate throughput achievable with a Gaussian codebook and nearest-neighbor decoding, for the case of linear precoding followed by oversampling {1-bit} DACs. This paper complements the analysis previously reported in~\cite{jacobsson17c} by generalizing it to oversampling multi-bit DACs.

Finally, nonlinear precoding algorithms for the 1-bit-DAC case have been  extended recently to OFDM and frequency-selective channels in~\cite{jacobsson17f, nedelcu17a, jedda17b}.

\subsection{Contributions}

We analyze the performance achievable in the massive MU-MIMO-OFDM downlink with linear precoding and finite-resolution DACs. In contrast to existing results~\cite{saxena16b, li17a, jacobsson17d, jacobsson16d, jedda16a, swindlehurst17a, castaneda17a, landau17a, li18a, jacobsson18b, shao18a, mezghani09c}, which consider symbol-rate DACs and single-carrier modulation over frequency-flat channels, we focus on oversampling DACs and OFDM-based transmission over frequency-selective channels. 
We focus on conventional linear precoders, namely MRT and ZF precoding. These precoders are of practical interest because they entail relatively low complexity compared to nonlinear precoders~\cite{jacobsson17f, nedelcu17a,jedda17b}.
Our main contributions can be summarized as~follows:
\begin{itemize}
\item Using Bussgang's theorem~\cite{bussgang52a, rowe82a}, we develop a lower bound on the information-theoretic sum-rate achievable with linear precoding and oversampling finite-resolution DACs. This lower bound can be achieved by Gaussian signaling and nearest-neighbor decoding at the UEs. We also show how to evaluate the lower bound  for the case of~$1$-bit~DACs using Van Vleck's arcsine~law~\cite{van-vleck66a}.
\item For the case of multi-bit DACs, we develop an approximate, yet accurate, model for the distortion caused by the DACs, which takes into account the inherent spatial and temporal correlation in the DAC distortion. This approach yields an accurate approximation for the SINDR at the UE side, which we use to evaluate the sum-rate lower bound and to derive an approximation of the uncoded BER achievable with QPSK. We also briefly discuss how this approximation can be used to characterize OOB emissions caused by~low-resolution~DACs.
\item We derive a simpler approximation for the DAC distortion, which treats this distortion as white (i.e., uncorrelated in both the spatial and the temporal domains). We show that such a crude model is accurate for medium-to-high resolution DACs and when the oversampling ratio~(OSR) is not too high, but is not sufficient to accurately describe the distortion caused by low-resolution~(e.g., 1 bit) DACs.
\item  We demonstrate through extensive numerical simulations that massive MU-MIMO-OFDM with low-resolution DACs enables excellent performance in terms of achievable rate and uncoded/coded BER. Specifically, we show that only a  few DAC bits (e.g., $3$--$4$ bits) are sufficient to approach the performance of systems that use infinite-resolution~DACs.
\end{itemize}

\subsection{Notation}

Lowercase and uppercase boldface letters represent column vectors and matrices, respectively. 
For a matrix $\bA$, we denote its complex conjugate, transpose, and Hermitian transpose by $\bA^*$,~$\bA^T$, and~$\bA^H$, respectively. The entry on the $k$th row and on the $\ell$th column of the matrix $\matA$ is denoted as $[\matA]_{k,\ell}$.
The $k$th entry of a vector $\veca$ is denoted as~$[\veca]_k$.
%
%
The trace and the main diagonal of~$\bA$ are~$\tr(\bA)$ and~$\text{diag}(\bA)$, respectively. The square matrix $\text{diag}(\veca)$ is diagonal with the elements of the vector $\veca$ along its main diagonal.
If~$\matA$ is an $M \times N$ matrix, then $\text{vec}(\matA)$ is the $MN$-dimensional vector obtained by stacking the columns (taken from left to right) of~$\matA$.
The $M\times M$ identity matrix, the $M \times N$ all-zeros matrix, and the $M \times N$ all-ones matrix are denoted by $\bI_M$, $\mathbf{0}_{M\times N}$, and~$\mathbf{1}_{M\times N}$, respectively.
The real and the imaginary parts of a complex vector~$\veca$ are denoted by $\Re\{\veca\}$ and $\Im\{\veca\}$, respectively. 
We use~$\vecnorm{\veca}$ to denote the $\ell_2$-norm of~$\veca$.
The Kronecker product of two matrices $\matA$ and $\matB$ is~$\matA \otimes \matB$.
For equally-sized matrices $\matA$ and $\matB$, we denote by $\matA \odot \matB$ the Hadamard (element-wise) product.
We use~$\text{sgn}(\cdot)$ to denote the signum function, which is applied entry-wise to vectors and defined as~$\text{sgn}(a)=1$ if~$a\ge0$ and~$\text{sgn}(a)=-1$ if~$a<0$.
We further use $\mathds{1}_{\setA}(a)$ to denote the indicator function, which is defined as~$\mathds{1}_{\setA}(a) = 1$ for $a \in \setA$ and $\mathds{1}_{\setA}(a) = 0$ for~$a \notin \setA$.
The multivariate complex-valued circularly-symmetric Gaussian distribution with covariance matrix~$\bK \in \opC^{N \times N}$ is denoted by~$\jpg(\bZero_{N \times 1},\bK)$. %
We use~$\Prob\lefto[\setE\right]$ to denote the probability of the event~$\setE$ and~$\Ex{}{x}$ to denote the expected value of~$x$.
The cumulative distribution function of the standard normal~distribution is~$\Phi(x) = \frac{1}{\sqrt{2\pi}}\int^{x}_{-\infty} \exp\lefto( -u^2/2 \right) \text{d}u$.
%

\subsection{Paper Outline}

The rest of the paper is organized as follows. In~\fref{sec:system_model}, we introduce the system model as well as the considered linear precoders, and we describe the operation of the DACs.
In \fref{sec:performance}, we review Bussgang's theorem, derive the SINDR at the UEs, and develop a lower bound on the rate achievable with Gaussian inputs and nearest-neighbor decoding. 
In~\fref{sec:quantization}, we derive exact and approximate expressions for the distortion caused by finite-resolution DACs. We also use these expressions to evaluate the achievable-rate lower bound from~\fref{sec:performance} and the uncoded BER achievable with QPSK. 
In \fref{sec:numerical}, we provide numerical simulations that demonstrate the accuracy of our analytical results. 
We conclude in~\fref{sec:conclusion}. 

\begin{figure*}[t]
\centering
 \includegraphics[width=.925\textwidth]{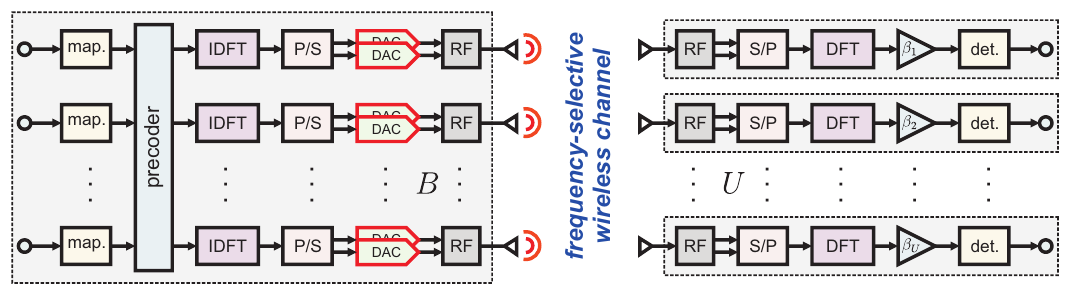}
 \caption{Overview of the massive MU-MIMO-OFDM downlink with low-resolution DACs at the BS. Left: A~$B$ antenna BS performs linear precoding and generates the per-antenna OFDM time-domain signals that are passed through low-resolution DACs (highlighted with red color). Right: $U$ single-antenna UEs perform independently OFDM demodulation and data detection. In the figure, ``map.''~and ``det.''~stand for mapper and detector, respectively.} 
  \label{fig:overview}
\end{figure*}


\section{System Model}
\label{sec:system_model}

We consider the single-cell downlink massive MU-MIMO-OFDM system depicted in \fref{fig:overview}.
The system consists of a BS with $B$ antennas that simultaneously serves $U$ single-antenna UEs in the same time-frequency resource using spatial multiplexing.
Our model includes finite-resolution DACs at the BS. Specifically, the in-phase and quadrature components of the time-domain per-antenna transmitted signal at the BS are generated using a pair of  finite resolution DACs.
Let $\setL = \{\ell_0, \ell_1, \dots , \ell_{L-1}\}$, where $\ell_0 <  \ell_1 < \dots < \ell_{L-2} < \ell_{L-1}$, denote the set of $L$ possible DAC outcomes, i.e., the set of possible amplitude (voltage) levels that are supported by the transcoder in the DACs.
The DACs have finite resolution, and, hence, the set $\setL$ has finite cardinality.
We refer to $L = \abs{\setL}$ and to $\log_2(L)$ as the number of DAC \emph{levels} and as the number of DAC \emph{bits} per real dimension, respectively. We assume all DACs at the BS to be the same. Hence, the set of complex-valued DAC outputs at each antenna is $\setX = \setL \times \setL$. 

The system operates over a wideband channel and OFDM is used to deal with the frequency selectiveness of the channel. Specifically, at the BS, the frequency-domain precoded vector is mapped to time domain by performing an inverse discrete Fourier transform~(IDFT) at each antenna element. At the UEs, the time-domain received signal is transformed back to frequency domain through a discrete Fourier transform~(DFT).  As we will show in \fref{sec:psd}, the nonlinearity introduced by the finite-resolution DACs will result in intercarrier interference~(ICI).


We assume that each OFDM symbol consists of $N$ time-domain samples. Let~$\Delta f$ be the subcarrier spacing and $f_s = N \Delta f$ be the sampling rate of the DACs.
We use the disjoint sets $\setS_d$ and $\setS_g$ to denote the set of subcarriers designated for the $S \le N$ data symbols (occupied subcarriers) and the set of $N-S$ guard subcarriers, respectively.
Let~$\vecs_k$~denote the $U$-dimensional data vector associated with the $k$th subcarrier~($k=0,1,\dots,N-1$).
We assume that $\vecs_k = \veczero_{U\times 1}$ for $k \in \setS_g$ and that $\Ex{}{\vecs_k\vecs_k^H} = \matI_U$ for~$k \in \setS_d$.
The case $S=N$ corresponds to symbol-rate-sampling DACs whereas $S< N$ corresponds to oversampling DACs. The OSR~is~$\osr = N/S$.

\subsection{Channel Input-Output Relation}

To isolate the  performance impact of low-resolution DACs, we assume in our analysis that all RF circuitry other than the DACs (e.g., local oscillators, mixers, and PAs) are ideal and that the UEs are equipped with infinite resolution ADCs.
We further assume that the sampling rate $f_s$ of the DACs at the BS equals the sampling rate of the ADCs at the UEs and that the system is perfectly synchronized.
Finally, we assume that the \emph{reconstruction stage} (see, e.g., \cite[Sec.~1.7]{maloberti07a}) of the DACs is an ideal low-pass filter with cutoff frequency $f_\text{cut} = f_s/2$ so that the spectrum of the DAC output is contained within~$[-f_s/2, f_s/2]$.\footnote{A more realistic reconstruction stage, specifically a zero-order hold filter followed by nonideal low-pass filter, is considered in~\cite{jacobsson17a}. For this case, the spectrum of the DAC output will not be contained within $[-f_s/2, f_s/2]$.}


\subsubsection{Time Domain}

With the above assumptions, the discrete-time baseband signal $\vecy_n \in \opC^U$ received at the $U$ UEs at time sample $n$ can be written~as 
\begin{IEEEeqnarray}{rCl} \label{eq:inout_time}
\vecy_n &=& \sum_{t=0}^{T-1} \matH_t \vecx_{n-t} + \vecw_n \IEEEeqnarraynumspace
\end{IEEEeqnarray}
for $n = 0, 1, \dots, N-1$.
Here, $\vecx_n \in \setX^B$ is the transmitted signal (i.e., the output of the DACs) at time sample $n$. The random vector $\vecw_n \sim \jpg(\veczero_{U \times 1},N_0\matI_U)$ models additive white Gaussian noise (AWGN) at the $u$th UE at time sample $n$. Here, $N_0$ is the power spectral density (PSD) of the~AWGN.
The matrix $\matH_t \in \opC^{U \times B}$ is the time-domain channel matrix associated with the $t$th tap of the frequency-selective channel ($t = 0, 1, \dots, T-1$, where~$T$ is the number of taps). We assume that this matrix has entries $\lefto[ \matH_t \right]_{u,b} \sim \jpg(0,T^{-1})$ and that the entries of $\lefto\{ \matH_t \right\}_{t=0}^{T-1}$ are independent and  remain constant over the duration of an OFDM symbol.
Note that these assumptions yield a spatially white frequency-selective Rayleigh-fading channel with uniform power-delay~profile.\footnote{Extensions to more general fading models that include, e.g., spatial correlation are immediate.}

A cyclic prefix of length $T-1$ is prepended to the transmitted signal and is later discarded at the UEs. We shall not explicitly prepend the cyclic prefix to $\lefto\{\vecx_n\right\}$ to keep notation compact. The cyclic prefix makes the channel matrix circulant and, hence, diagonalizable in the infinite-resolution case through an IDFT and a DFT at the BS and at the UEs, respectively. The transmitted signal $\lefto\{ \vecx_n \right\}$ satisfies the average power constraint
\begin{IEEEeqnarray}{rCl} \label{eq:powerconstraint}
	\Ex{}{ \sum_{n=0}^{N-1}\vecnorm{\vecx_n}^2} \le PS.
\end{IEEEeqnarray}
Here, $0 < P < \infty$ is the average transmit power at the BS. We define $\snr = P / N_0$ as the~SNR.

\subsubsection{Frequency Domain}

Let $\matX = \lefto[ \vecx_0, \vecx_1, \dots, \vecx_{N-1} \right] \in \setX^{B\times N}$ and $\matY = \lefto[ \vecy_0, \vecy_1,  \dots, \vecy_{N-1} \right]\in \complexset^{U\times N}$ be the time-domain transmitted and received matrices over the~$N$ time samples, respectively. 
Furthermore, let $\widehat\matX  = \matX\matF_N^T $ and $\widehat\matY  = \matY\matF_N^T$ be the corresponding frequency-domain matrices. 
Here,~$\matF_N$ stands for the $N \times N$ DFT matrix, which satisfies $\matF_N\matF_N^H = \matI_N$.
Finally,~let $\widehat\matH_{k} = \sum_{t=0}^{T-1} \matH_t \exp\lefto( - jk\frac{2\pi}{N}t \right)$ for $k=0,1,\dots, N-1$, be the $U \times B$ frequency-domain channel matrix associated with the $k$th~subcarrier.
After discarding the cyclic prefix, we can write the frequency-domain input-output relation at the $k$th subcarrier~as
\begin{IEEEeqnarray}{rCl} \label{eq:inout_frequency}
	\hat\vecy_{k}	&=& \widehat\matH_{k}\hat\vecx_{k} + \hat\vecw_{k}.
\end{IEEEeqnarray}
Here, $\hat\vecx_{k}$ and $\hat\vecy_{k}$ are the $k$th column of $\widehat\matX$ and $\widehat\matY$, respectively.
Furthermore,~$\hat\vecw_{k} \distas \jpg(\veczero_{U \times 1}, N_0 \matI_U)$ is  the $k$th column of the matrix $\widehat\matW = \matW\matF_N^T$, where $\matW = \lefto[ \vecw_0, \vecw_1, \dots, \vecw_{N-1} \right]$.
In the following two subsections, we shall relate the frequency-domain representation~$\hat\vecx_k$ of the DAC output to the precoded transmitted symbols.

\subsection{Uniform Quantization}
\label{sec:uniform}

For the discrete-time system model~\eqref{eq:inout_time} considered in this paper (recall that this model assumes that the reconstruction stage in the DACs is an ideal low-pass filter), each DAC can be modeled simply as a \emph{quantizer},\footnote{We assume that the baseband processing unit at the BS uses floating-point arithmetic with infinite word length. The impact of having finite word lengths in massive MU-MIMO baseband processing has been investigated in~\cite{gunnarsson17a}.} i.e., a nonlinear device that maps a continuous-amplitude signal to a set of discrete numbers~\cite{gray98a}.
We characterize the quantizer by the set $\setL = \{ \ell_0, \ell_1, \dots. \ell_{L-1}\}$ of $L$ quantization labels and the set $\setT = \{\tau_0, \tau_1, \dots, \tau_{L} \}$, where $-\infty = \tau_0 < \tau_1 < \dots < \tau_{L-1} <\tau_{L} = \infty$, of~$L+1$ quantization thresholds.
We use the \emph{quantization function} $\quantize(\cdot) : \opC^B \rightarrow \setX^B$, which is applied entry-wise to a vector, to describe the joint operation of the $2B$ DACs at the~BS. Let $\vecz_n \in \opC^B$ denote the time-domain precoded vector at time sample~$n$.\footnote{In \fref{sec:linear_precoding}, we shall describe how the time-domain precoded vectors~$\{ \vecz_n \}$ for $n = 0,1,\dots,N-1$ are obtained from the data symbols~$\{\vecs_k\}$ for~$k \in \setS_d$.} Also, let $z_{b,n} = \lefto[\vecz_n\right]_b$ and $x_{b,n} = \lefto[\vecx_n\right]_b$.~Then, 
\begin{IEEEeqnarray}{rCl} 
 x_{b,n} &=& \quantize(z_{b,n}) \\[2pt]
&=& \sum_{i=0}^{L-1} \ell_i \mathds{1}_{[\tau_i,\tau_{i+1})}\lefto( z_{b,n}^R \right) + j \sum_{i=0}^{L-1} \ell_i \mathds{1}_{[\tau_i,\tau_{i+1})}\lefto( z_{b,n}^I  \right) \IEEEeqnarraynumspace \label{eq:quantizer}
\end{IEEEeqnarray}
where $z_{b,n}^R = \Re\lefto\{z_{b,n}\right\}$ and $z_{b,n}^I = \Im\lefto\{z_{b,n}\right\}$.
For simplicity, we shall model the DACs as symmetric {uniform} quantizers (the labels $\{\ell_i\}$ are equispaced and symmetric around zero) with step size $\Delta$. For symmetric uniform quantizers, the quantization thresholds are $\tau_i = \Delta \lefto( i - \frac{L}{2}\right)$ for $i=1,2,\dots,L-1$ and $\tau_0 = -\infty$, $\tau_L = \infty$. 
Furthermore, the quantization labels are $\ell_i = \alpha\Delta \lefto( i - \frac{L}{2} + \frac{1}{2}\right)$ for $i=0,1,\dots,L-1$. Note that the quantization labels are scaled by a constant~$\alpha$ to ensure that the transmit power constraint~\eqref{eq:powerconstraint} is satisfied.
If $L$ is odd, then the quantizer has a label at zero; we shall refer to such quantizers as a \emph{midtread} quantizers. If $L$ is even, then the quantizer has a threshold at zero; we shall refer to such quantizers as a \emph{midrise}~quantizers.

The choice of the step size $\Delta$ determines the amount of distortion caused by the DACs. If $\Delta$ is too small, then there will be significant \emph{overload} distortion (clipping or saturation); if $\Delta$ is too large, then there will be significant \emph{granular} distortion.  
Specifically, let $A_\text{clip} = {L\,\Delta}/{2}$ be the \emph{clipping level} of the uniform quantizer.
The overload distortion is the error $\alpha^{-1}\quantize(z) - z$ occurring if $\abs{z} > A_\text{clip}$; the granular distortion is the error $\alpha^{-1}\quantize(z) - z$ occurring if $\abs{z} \le A_\text{clip}$. 
It will turn out important for our analysis to keep the overload distortion negligible compared to the granular distortion. Therefore, we will chose the step size~$\Delta$ such that~the probability of the event~$\abs{z} > A_\text{clip}$ is ``small'' (we will discuss this important aspect in Sections~\ref{sec:quantizer_rounding}~and~\ref{sec:numerical}).
In the extreme case of 1-bit DACs, the quantization function~\eqref{eq:quantizer} reduces to
\begin{IEEEeqnarray}{rCl} \label{eq:quantizer_1bit}
x_{b,n} = \quantize\lefto(z_{b,n}\right) = \frac{\alpha\Delta}{2}\big( \sign(z_{b,n}^R) + j\sign( z_{b,n}^I)\big). \IEEEeqnarraynumspace 
\end{IEEEeqnarray}
Here, by setting $\alpha = \sqrt{2P/(\Delta^2 \osr B)}$, we ensure that the power constraint~\eqref{eq:powerconstraint} is satisfied with equality (recall that $\xi = N/S$ is the OSR).
%
%

\subsection{Linear Precoding}
\label{sec:linear_precoding}

At the BS, the data symbols for the $U$ UEs are mapped to the antenna array by a precoder. We focus, in this paper, only on linear precoders because of their low computational complexity and because a performance analysis is analytically~tractable. 
%

We assume that the BS has access to perfect CSI,\footnote{In~\fref{sec:csi_error}, we will relax this assumption by investigating the impact of imperfect CSI on BER performance in the MU-MIMO-OFDM downlink in the presence of low-resolution DACs.} i.e., it has perfect knowledge of the realizations of the frequency-domain channel matrices $\big\{\widehat\matH_{k}\big\}$ for $k \in \setS_d$. With linear precoding, the transmitted vector~$\vecx_n$ can be written as
\begin{IEEEeqnarray}{rCl}
\vecx_n &=& \quantize\lefto( \vecz_n \right)
\end{IEEEeqnarray}
where $\vecz_n \in \opC^{B}$ denotes the time-domain precoded vector at time sample $n$, which is obtained from the data symbols $\lefto\{ \vecs_k \right\}$ for $k \in \setS_d$ as
\begin{IEEEeqnarray}{rCl} \label{eq:vecz_time}
\vecz_n &=& \frac{1}{\sqrt{N}}\sum_{k \in \setS_d} \widehat\matP_{k}\vecs_k \exp\lefto(jk\frac{2\pi}{N}n\right)
\end{IEEEeqnarray}
for $n = 0, 1, \dots, N-1$. In words, the data symbols on the $k$th subcarrier ($k \in \setS_d$) are multiplied in the frequency domain with per-subcarrier  \emph{precoding matrices} $\widehat\matP_{k} \in \opC^{B \times U}$. The resulting frequency-domain precoded vector is then mapped to time domain by an IDFT. We use the convention that~$\widehat\matP_{k} = \matzero_{B \times U}$ for $k \in \setS_g$. 
In what follows, we will focus on two linear precoders that are commonly studied in the infinite-resolution case, namely MRT and ZF~precoding.

%

\subsubsection{MRT Precoding} 

The MRT precoder maximizes the power directed towards each UE, ignoring MU interference. The MRT precoding matrices are given~by
\begin{IEEEeqnarray}{rCl} \label{eq:Pmrt}
\widehat\matP_{k}^\text{MRT} &=& \frac{1}{\beta_\text{MRT}B} \widehat\matH_{k}^H
\end{IEEEeqnarray}
for $k \in \setS_d$, where
\begin{IEEEeqnarray}{rCl}
\beta^\text{MRT} = \sqrt{\frac{1}{PSB^2}\sum_{k \in \setS_d} \tr\!\big(\widehat\matH_{k} \widehat\matH_{k}^H\big)}
\end{IEEEeqnarray}
ensures that the power constraint~\eqref{eq:powerconstraint} is satisfied (in the infinite-resolution case)

\subsubsection{ZF Precoding} 

With ZF precoding, the BS nulls (in the infinite-resolution case) the MU interference by choosing the pseudo-inverse of the channel matrix as precoding matrix. 
For $k \in \setS_d$,  the ZF precoding matrices are given by
\begin{IEEEeqnarray}{rCl} \label{eq:Pzf}
	\widehat\matP^\text{ZF}_{k} &=& \frac{1}{\beta^\text{ZF}} \widehat\matH_{k}^H \lefto( \widehat\matH_{k} \widehat\matH_{k}^H\right)^{-1}
\end{IEEEeqnarray}
for $k \in \setS_d$, where
\begin{IEEEeqnarray}{rCl} 
\beta^\text{ZF} = \sqrt{\frac{1}{PS} \sum_{k \in \setS_d} \tr\big( \widehat\matH_{k} \widehat\matH_{k}^H\big)^{-1}}
\end{IEEEeqnarray}
ensures that the power constraint~\eqref{eq:powerconstraint} is satisfied (in the infinite-resolution case).

\section{Performance Analysis}
\label{sec:performance}

In the infinite-resolution case, the frequency-domain received signal $\hat\vecy_{k}$ in \eqref{eq:inout_frequency} can be written~as
\begin{IEEEeqnarray}{rCl} \label{eq:nicelinear}
	\hat\vecy_{k} &=& \widehat\matH_{k}\widehat\matP_{k} \vecs_k + \hat\vecw_k	.
\end{IEEEeqnarray}
In words, the received signal on subcarrier $k \in \setS_d$ depends only on $\vecs_k$ and not on the data symbols transmitted on other subcarriers. Hence, each subcarrier can be analyzed separately.
In the finite-resolution-DAC case, however, the nonlinearity introduced by the finite-resolution DACs through~\eqref{eq:quantizer} makes the received signal on one subcarrier depend, in general, on~the~data symbols transmitted on all other subcarriers.
To enable a performance analysis, it is convenient to write the frequency-domain received signal $\widehat\matY$ in vectorized form $\hat\vecy = \text{vec}(\widehat\matY) \in \opC^{UN}$~as
\begin{IEEEeqnarray}{rCl} \label{eq:inout_vectorized_temp} 
	\hat\vecy &=& \widehat\matH  \hat\vecx + \hat\vecw.
\end{IEEEeqnarray}
Here, $\hat\vecx = \text{vec} (\widehat\matX) \in \opC^{BN}$, $\hat\vecw = \text{vec}(\widehat\matW) \in \opC^{UN}$, and $\widehat\matH$ is the $U N \times B N$ block-diagonal matrix that has the matrices $\widehat\matH_{0},\widehat\matH_{1},\dots,\widehat\matH_{N-1}$ on its main diagonal. Next, we rewrite~$\hat\vecx$~as
\begin{IEEEeqnarray}{rCl} \label{eq:kronecker}
	\hat\vecx &=& \text{vec}(\widehat\matX)	= \text{vec}(\matX\matF_N^T) = \lefto( \matF_N \otimes \matI_B \right)\vecx \IEEEeqnarraynumspace
\end{IEEEeqnarray}
where $\vecx = \text{vec}(\matX) = \quantize(\vecz) \in \setX^{BN}$, with~$\vecz = \text{vec}\lefto( \matZ \right) \in \opC^{BN}$ and $\matZ = \lefto[ \vecz_0, \vecz_1, \dots, \vecz_{N-1}\right]$. 
Now let~$\widehat\matP \in \opC^{B N \times U N}$ denote the block-diagonal matrix that contains the matrices $\widehat\matP_{0}, \widehat\matP_{1}, \dots, \widehat\matP_{N-1}$ on its main diagonal.
With these definitions, we can compactly write the discrete-time precoded vector~$\vecz$~as
\begin{IEEEeqnarray}{rCl} \label{eq:vecz_precoded_vectorized}
\vecz = \lefto( \matF_N^H  \otimes \matI_B \right) \widehat\matP \vecs	
\end{IEEEeqnarray}
where $\vecs = \text{vec}(\matS)$ and $\matS = \lefto[ \vecs_0, \vecs_1, \dots, \vecs_{N-1}\right]$. 
Now, using~\eqref{eq:kronecker} and~\eqref{eq:vecz_precoded_vectorized} in~\eqref{eq:inout_vectorized_temp},  we obtain
\begin{IEEEeqnarray}{rCl} 
	\hat\vecy 
&=& \widehat\matH \lefto( \matF_N \otimes \matI_B \right) \vecx + \hat\vecw \\
&=& \widehat\matH \lefto( \matF_N \otimes \matI_B \right) \quantize\lefto( \lefto( \matF_N^H  \otimes \matI_B \right) \widehat\matP \vecs\right) + \hat\vecw. \IEEEeqnarraynumspace \label{eq:inout_vectorized}
\end{IEEEeqnarray}
Comparing~\eqref{eq:nicelinear} and \eqref{eq:inout_vectorized}, we see how the nonlinearity introduced by $\quantize(\cdot)$ complicates the input-output relation, thus, preventing a straightforward evaluation of the performance of the MU-MIMO-OFDM downlink system. We next use Bussgang's theorem~\cite{bussgang52a} to decompose~\eqref{eq:inout_vectorized} into a form that enables an analytical performance analysis and that captures the ICI caused by the~DACs.

\subsection{Decomposition Using Bussgang's Theorem}
\label{sec:bussgang}

The finite-resolution DACs introduce a quantization error $\vece \in \opC^{BN}$,
\begin{IEEEeqnarray}{rCl} \label{eq:error}
\vece &=& \alpha^{-1}\vecx - \vecz = \alpha^{-1}\quantize\lefto(\vecz\right) - \vecz.
\end{IEEEeqnarray}
The multiplicaction by $\alpha^{-1}$ is necessary because  we scaled the quantization labels by $\alpha$ to satisfy the power constraint~\eqref{eq:powerconstraint}.
Note that $\vece$ is \emph{correlated} with the quantizer input $\vecz$. 
%
%
For Gaussian inputs, Bussgang's theorem~\cite{bussgang52a} allows us to decompose $\quantize(\vecz)$ into two components: a linear function in $\vecz$ and a distortion that is \emph{uncorrelated} with~$\vecz$. Specifically, assume that $\vecs_k\distas\jpg( \veczero_{U \times 1}, \matI_{U})$ for all $k \in \setS_d$, and that the $\{ \vecs_k \}$ for $k \in \setS_d$ are independent. 
Let $x_m$ and $z_m$ be the $m$th element of $\vecx$ and $\vecz$, respectively. According to Bussgang's theorem~\cite{bussgang52a}, it~holds~that 
\begin{IEEEeqnarray}{rCl} \label{eq:bussgang_scalar}
	\Ex{}{x_m z_{n}^*} &=&  g_m \Ex{}{z_mz_{n}^*}
\end{IEEEeqnarray}
where $g_m = \sigma_m^{-2} \Ex{}{x_m z_m^*}$ and $\sigma_m^{2} = \Ex{}{|z_m|^2}$. Define now $\matC_{\vecx\vecz} = \Ex{}{\vecx\vecz^H} \in \opC^{BN \times BN}$ and $\matC_\vecz = \Ex{}{\vecz\vecz^H} \in \opC^{BN \times BN}$. It follows from~\eqref{eq:bussgang_scalar} that  $\matC_{\vecx\vecz} = \matG \matC_\vecz$ with $\matG = \text{diag}\lefto([g_1, g_2, \dots, g_{BN}]^T \right)$. Consequently, the transmitted time-domain vector~$\vecx$ can be written as
\begin{IEEEeqnarray}{rCl} \label{eq:inout_distortion} 
	\vecx &=& \quantize(\vecz) = \matG \vecz + \vecd	
\end{IEEEeqnarray}
where the distortion $\vecd \in \opC^{BN}$ is uncorrelated with $\vecz$, i.e., $\Ex{}{\vecz\vecd^H} = \matzero_{BN \times BN}$.~Indeed,
\begin{IEEEeqnarray}{rCl} 
\Ex{}{\vecd\vecz^H} 
&=& 	\Ex{}{(\vecx - \matG\vecz)\vecz^H} \\
&=& \matC_{\vecx\vecz} - \matG\matC_\vecz = \matzero_{BN \times BN}.
\end{IEEEeqnarray}
Here, we used that $\matC_{\vecx\vecz} = \matG \matC_\vecz$. It turns out that the real-valued diagonal matrix $\matG$ can be evaluated in closed form. Indeed, by following the same steps as in \cite[App.~A]{jacobsson17d}, we find~that
\begin{IEEEeqnarray}{rCl} \label{eq:gainmatrix}
	\matG = \matI_N \otimes \text{diag}\lefto( \vecg \right)
\end{IEEEeqnarray}
where $\text{diag}\lefto( \vecg \right) \in \opR^{B \times B}$ is given by 
\begin{IEEEeqnarray}{rCl} \label{eq:gainmatrix_uniform}
\text{diag}\lefto( \vecg \right) &=& \frac{\alpha\Delta}{\sqrt{\pi}} \, \text{diag}\lefto( \frac{1}{N} \sum_{k \in \setS_d} \widehat\matP_{k}\widehat\matP_{k}^H \right)^{-1/2} \nonumber \\
&& \times \sum_{i=1}^{L-1} \exp\!\Bigg(-\Delta^2\lefto( i - \frac{L}{2} \right)^2 \nonumber\\
&& \times \text{diag}\lefto( \frac{1}{N} \sum_{k \in \setS_d} \widehat\matP_{k}\widehat\matP_{k}^H \right)^{-1} \Bigg). \IEEEeqnarraynumspace
\end{IEEEeqnarray}
For the special case of 1-bit DACs, \eqref{eq:gainmatrix_uniform} reduces to
\begin{IEEEeqnarray}{rCl} \label{eq:gainmatrix_1bit}
	\text{diag}(\vecg) &=& \sqrt{\frac{2P}{\pi \osr B}	} \, \text{diag}\lefto( \frac{1}{N}  \sum_{k \in \setS_d}  \widehat\matP_{k} \widehat\matP_{k}^H \right)^{\!-1/2}\!\!. \IEEEeqnarraynumspace
\end{IEEEeqnarray}
Inserting~\eqref{eq:inout_distortion} into \eqref{eq:inout_vectorized}, we obtain
\begin{IEEEeqnarray}{rCl}
\hat\vecy 
&=& \widehat\matH \lefto( \matF_N \otimes \matI_B \right) \lefto(\matG \lefto( \matF_N^H  \otimes \matI_B \right) \widehat\matP  \vecs +  \vecd\right) + \hat\vecw \\
&=& \widehat\matH \matG \widehat\matP  \vecs + \widehat\matH \lefto( \matF_N \otimes \matI_B \right) \vecd + \hat\vecw \label{eq:received_final}
\end{IEEEeqnarray}
where the last step holds because $\lefto( \matF_N \otimes \matI_B \right) \matG \lefto( \matF_N^H  \otimes \matI_B \right) = \matG$, as a consequence of~\eqref{eq:gainmatrix}.


\subsection{Achievable Sum-Rate with Gaussian Inputs}
\label{sec:achievable}

Let $\hat{y}_{u,k} = \lefto[ \hat\vecy_{k} \right]_u$ denote the  received signal on the $k$th subcarrier and at the $u$th UE. 
It follows from~\eqref{eq:received_final}~that
\begin{IEEEeqnarray}{rCl} 
\hat{y}_{u,k} &=&	\lefto[\widehat\matH_{k\,} \text{diag}\lefto(\vecg\right) \widehat\matP_{k}\right]_{u,u}  \! s_{u,k} \nonumber\\
&& + {\sum_{v \neq u}} \lefto[\widehat\matH_{k\,} \text{diag}\lefto(\vecg\right) \widehat\matP_{k}\right]_{u,v}  \! s_{v,k} \nonumber\\
&& + [\widehat\matH \lefto( \matF_N \otimes \matI_B \right) \vecd]_ {u+kU} + \hat{w}_{u,k} \IEEEeqnarraynumspace \label{eq:decomposition}
\end{IEEEeqnarray}
where~$s_{u,k} = [\vecs_k]_u$ and~$\hat{w}_{u,k} = [\hat\vecw_k]_u$. The first term on the right-hand side (RHS) of~\eqref{eq:decomposition} corresponds to the desired signal; the second term captures the MU interference; the third term describes the distortion introduced by the DACs; the fourth term represents the~AWGN.

Let now $\sindr_{u,k}(\widehat\matH)$ be the SINDR on the $k$th subcarrier for the~$u$th~UE.
Using~\eqref{eq:decomposition} and assuming that $\vecs_k \sim \jpg(\veczero_{U \times 1}, \matI_U)$ for $k \in \setS_d$, we can express $\sindr_{u,k}(\widehat\matH)$ as
\begin{IEEEeqnarray}{rCl} \label{eq:sindr_mimo}
\sindr_{u,k}(\widehat\matH) &=& \frac{\lefto[\big\lvert\widehat\matH_{k\,} \text{diag}\lefto(\vecg\right) \widehat\matP_{k}\big\rvert^2\right]_{u,u}}{\sum\limits_{v \neq u}\!\big[\big\lvert\widehat\matH_{k}\text{diag}\lefto( \vecg \right) \! \widehat\matP_{k}\big\rvert^2\big]_{u,v} \! + D_{u,k}(\widehat\matH)+ N_0}  \IEEEeqnarraynumspace
\end{IEEEeqnarray}
%
%
where
\begin{IEEEeqnarray}{rCl} 
\IEEEeqnarraymulticol{3}{l}{
D_{u,k}(\widehat\matH)
} \nonumber\\ \ &=& \lefto[ \widehat\matH \lefto( \matF_N  \otimes \matI_B \right) \matC_{\vecd} \lefto( \matF_N^H  \otimes \matI_B \right) \widehat\matH^H\right]_{u + kU, u + kU}. \IEEEeqnarraynumspace
\end{IEEEeqnarray}
Here, $\matC_\vecd = \Ex{}{\vecd\vecd^H} \in \opC^{BN \times BN}$ is the covariance of the distortion $\vecd$. For the case of symbol-rate-sampling DACs and single-carrier transmission ($S=N=1$) over a frequency-flat channel ($T=1$), the SINDR in~\eqref{eq:sindr_mimo} simplifies to the SINDR reported in~\cite[Eq.~(26)]{jacobsson17d}.

Due to the nonlinearity introduced by the DACs, the distortion $\vecd$ is not Gaussian distributed, which makes it challenging to compute exactly the achievable rate. It is, however, possible to derive a lower bound the achievable rate using the so-called ``auxiliary-channel lower bound''~\cite[p.~3503]{arnold06a}.
Through standard manipulations of the mutual information~(see, e.g., \cite[Sec.~III-D]{jacobsson17d}), we obtain the following lower bound on the ergodic sum rate\footnote{We assume that coding can be performed over sufficiently many independent realization of $\{\matH_t\}$ for ${t = 0,1,\dots,T-1}$.} that is explicit in the SINDR~\eqref{eq:sindr_mimo}
\begin{IEEEeqnarray}{rCl} \label{eq:rate}
	R _\text{sum} &=&  \frac{1}{S} \Ex{}{\sum_{u=1}^U\sum_{k \in \setS_d} \log_2\lefto( 1 + \sindr_{u,k}(\widehat\matH)\right)}
\end{IEEEeqnarray}
where the expectation is over the channel matrix $\widehat\matH$.
It follows from  a generalized mutual information analysis, similar to the one reported in~\cite{lapidoth96b, zhang12a}, that the lower bound in~\eqref{eq:rate} corresponds to the ergodic sum rate achievable using a Gaussian codebook and a mismatched scaled nearest-neighbor decoder at the UEs under the assumption that the channel gains\footnote{These are the per-subcarrier scaling factors used in the scaled nearest-neighbor decoding rule.} $\big[\widehat\matH_{k\,} \mathrm{diag}\lefto(\vecg\right) \widehat\matP_{k}\big]_{u,u}$ for $k \in \setS_d$ are perfectly known to the $u$th UE, $u = 1, 2,  \dots, U$.

\section{Exact and Approximate Distortion Models}
\label{sec:quantization}

We next tackle the problem of evaluating the covariance matrix $\matC_\vecd$ of the distortion $\vecd$, which is required to compute~\eqref{eq:sindr_mimo} and, hence, the achievable rate~\eqref{eq:rate}. As in the previous section, we assume Gaussian signaling, i.e., that $\vecs_k \sim \jpg(\veczero_{U \times 1}, \matI_U)$ for $k \in \setS_d$. This implies that~$\vecz\distas\jpg(\veczero_{BN \times 1}, \matC_\vecz)$ where
\begin{IEEEeqnarray}{rCl} \label{eq:Czz}
  \matC_\vecz = \lefto( \matF_N^H  \otimes \matI_B \right) \widehat\matP \widehat\matP^H \lefto( \matF_N \otimes \matI_B \right).
\end{IEEEeqnarray}
It follows from \eqref{eq:inout_distortion} that
\begin{IEEEeqnarray}{rCl} \label{eq:Cdd}
\matC_\vecd 
&=& \Ex{}{\vecd\vecd^H} 
= \matC_\vecx - \matG \matC_\vecz \matG.
\end{IEEEeqnarray}
Here, we used that $\matG^H = \matG$ since $\matG$ is a real-valued diagonal matrix, and that $\matC_{\vecx\vecz} = \matG\matC_\vecz$.
Hence, to evaluate $\matC_\vecd$, one has to compute the covariance $\matC_\vecx = \Ex{}{\vecx\vecx^H} \in \opC^{BN \times BN}$ of the DAC output.
We next discuss how to evaluate $\matC_\vecx$.

\subsection{Computation of $\matC_\vecx$}
\label{sec:quantizer_exact}

Let $x_m = [\vecx]_m$ and $x_n = [\vecx]_n$. 
We can write the entry on the $m$th row and $n$th column of~$\matC_\vecx$ as $\Ex{}{x_m x_n^*}$.
Let now $x_m^R = \Re\{ x_m \}$ and $x_m^I = \Im\{ x_m \}$ denote the real and imaginary components of $x_m$, respectively. 
Since the input to the DACs is a circularly-symmetric Gaussian random variable, $\Ex{}{x_m x_n^*}$ can be as expanded as follows:
\begin{IEEEeqnarray}{rCl} 
\Ex{}{x_m x_n^*} 
&=& 2\Big(\Ex{}{x_m^R x_n^R} + j\Ex{}{x_m^I x_n^R}\Big). \IEEEeqnarraynumspace \label{eq:covariance_xx}
\end{IEEEeqnarray}
Let now $z_m$ be the $m$th element of $\vecz$ and let $z_m^R = \Re\{ z_m \}$. For the case $n = m$, for which it holds that $\Ex{}{x_m^I x_m^R} = 0$,~\eqref{eq:covariance_xx} reduces to
\begin{IEEEeqnarray}{rCl}
\Ex{}{\abs{x_m}^2} 
&=& 2 \sum_{i=0}^{L-1} \ell_i^2 \Prob[x_m^R = \ell_i] \\
&=& 2 \sum_{i=0}^{L-1} \ell_i^2 \Prob[\tau_{i} \le z_m^R < \tau_{i+1}] \\
&=& 2 \sum_{i=0}^{L-1} \ell_i^2 \lefto( \Phi\lefto(\frac{\sqrt{2}\,\tau_{i+1}}{\sigma_{m}}\right) - \Phi\lefto( \frac{\sqrt{2}\,\tau_{i}}{\sigma_{m}}\right) \right) \IEEEeqnarraynumspace \label{eq:xm2_temp} \\
&=& \frac{\alpha^2\Delta^2}{2}\lefto(L-1\right)^2 \nonumber\\
&& - 4\alpha^2\Delta^2\sum_{i=1}^{L-1}\lefto(i-\frac{L}{2}\right)\Phi\lefto(\frac{\sqrt{2}}{\sigma_m}\lefto(i-\frac{L}{2}\right)\right) \IEEEeqnarraynumspace \label{eq:xm2}
\end{IEEEeqnarray}
where \eqref{eq:xm2_temp} follows because $z_m \sim \jpg(0, \sigma_m^2)$ with $\sigma_{m}^2 = \Ex{}{|z_m|^2} = [\matC_\vecz]_{m,m}$. To derive~\eqref{eq:xm2}, we used that, for the uniform quantizer considered in this work, we have $\ell_{i+1} = \ell_i + \alpha\Delta$ for $i \in \{ 0, 1, \dots, L-2 \}$.
For the case $m \neq n$, the expectation $\Ex{}{x_m^C x_n^R}$, where $C \in \{ R, I \}$, can be written~as
\begin{IEEEeqnarray}{rCl}
\IEEEeqnarraymulticol{3}{l}{
\Ex{}{x_m^C x_n^R} 
} \nonumber\\
&=& \sum_{a=0}^{L-1} \sum_{b=0}^{L-1} \ell_{a}\ell_{b} \Prob[x_m^C = \ell_{a}, x_n^R = \ell_{b}] \label{eq:covariance_xRxR}\\
&=& \sum_{a=0}^{L-1} \sum_{b=0}^{L-1} \ell_{a}\ell_{b} \Prob[\tau_{a} \le z_m^C < \tau_{a+1} , \tau_{b} \le z_n^R < \tau_{b+1}]. \label{eq:integral_no_closed_form} \IEEEeqnarraynumspace 
\end{IEEEeqnarray}
Unfortunately,~\eqref{eq:integral_no_closed_form} does not  have a known closed-form expression and hence, has to be evaluated using numerical methods~(cf. \cite{li17e}). 
One exception is the special case of 1-bit DACs ($L=2$), for which it holds~that
\begin{IEEEeqnarray}{rCl} 
\IEEEeqnarraymulticol{3}{l}{
\Ex{}{x_m x_n^*}
} \nonumber\\
&=& \frac{2 P }{\pi \osr B} \lefto(\arcsin\!\bigg(\frac{\sigma_{m,n}^R}{\sigma_m\sigma_n}\bigg) + j \arcsin\!\bigg(\frac{\sigma_{m,n}^I}{\sigma_m\sigma_n}\bigg)\right). \IEEEeqnarraynumspace \label{eq:arcsine_siso}
\end{IEEEeqnarray}
Here, we have defined $\sigma_{m,n}^R = \Re\lefto\{[\matC_\vecz]_{m,n}\right\}$ and $\sigma_{m,n}^I = \Im\lefto\{[\matC_\vecz]_{m,n}\right\}$.
%
%
This well-known result, reported first by Van Vleck and Middleton~\cite{van-vleck66a}, is commonly referred to as the  \emph{arcsine law}. Writing \eqref{eq:arcsine_siso} in matrix form, we obtain, for the 1-bit-DAC case,
\begin{IEEEeqnarray}{rCl} \label{eq:arcsine_mimo}
	\matC_{\vecx}
	&=&\frac{2 P }{\pi \osr B}\lefto(\arcsin\lefto( \text{diag}(\matC_{\vecz})^{-\frac{1}{2}} \, \Re\{ \matC_{\vecz} \} \, \text{diag}(\matC_{\vecz})^{-\frac{1}{2}}\right) \right. \nonumber\\ 
	&&\lefto.+j\arcsin\lefto( \text{diag}(\matC_{\vecz})^{-\frac{1}{2}} \, \Im\{ \matC_{\vecz} \} \, \text{diag}(\matC_{\vecz})^{-\frac{1}{2}}\right) \right). \ \IEEEeqnarraynumspace
\end{IEEEeqnarray}
By inserting~\eqref{eq:arcsine_mimo} into \eqref{eq:Cdd} we find the desired covariance matrix $\matC_\vecd$, which allow us to compute the SINDR~\eqref{eq:sindr_mimo} and the ergodic sum rate~\eqref{eq:rate} for the 1-bit-DAC case.

To evaluate $\matC_\vecd$ in~\eqref{eq:Cdd} for the case~$L>2$, one has to compute~\eqref{eq:integral_no_closed_form} using numerical integration, which is time consuming and offers limited insights.\footnote{For the uplink case, focusing the quantization distortion caused by symbol-rate-sampling ADCs, the required covariance matrix was found by means of numerical integration in~\cite{li17e}. This approach, however, is not practical in our setup as the number of entries in $\matC_\vecx$ scales quadratically in $BN$, where $B$ could be in the order of hundreds and $N$ could be in the order of~thousands.} 
In what follows, we shall present two closed-form approximations for $\matC_\vecx$, which trade accuracy for complexity in distinct ways.
%

%

%



\subsection{Rounding Approximation}
\label{sec:quantizer_rounding}
 
\begin{figure}[!t]
\centering
\subfloat[$\alpha^{-1}\quantize(z)$: uniform midrise quantizer with $L = 6$ levels.]{\includegraphics[width = .95\columnwidth]{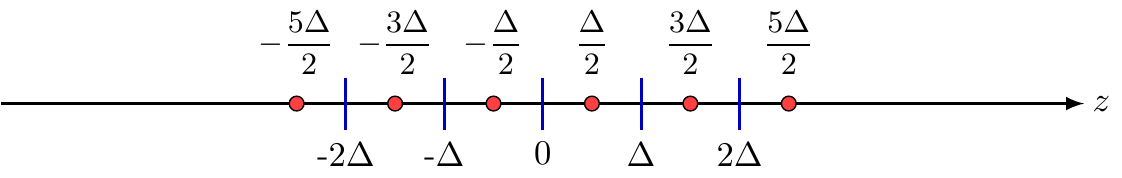}} \\
\subfloat[$\rounding(z)$: uniform midrise quantizer with an infinite number of levels.]{\includegraphics[width = .95\columnwidth]{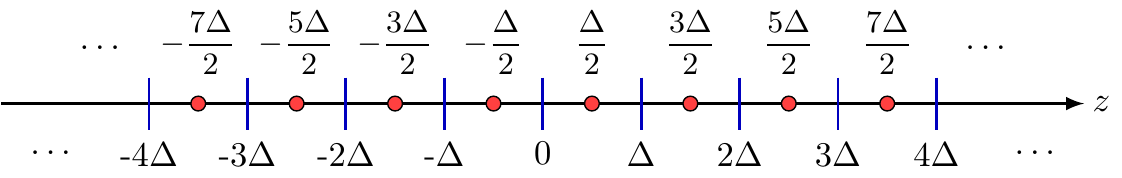}}
\caption{Comparison between the quantization levels and thresholds associated with the rules $\alpha^{-1}\quantize(z)$ and $\rounding(z)$; the red circles correspond to quantization labels, and the blue vertical lines correspond to the quantization thresholds.}
\label{fig:quantizer_and_rounder}
\end{figure}

First, we present a \emph{rounding approximation} $\matC_\vecd^\text{round}$ of $\matC_\vecd$.
This approximation takes into account the correlation between the entries of $\vecd$ and  turns out to be accurate as long as the step size $\Delta$ of the DACs is set so that the overload distortion is negligible compared to the granular~distortion.
To derive this approximation, we start by noting from~\eqref{eq:error} that $\matC_{\vecx}$ can be written~as
\begin{IEEEeqnarray}{rCl} \label{eq:outcovariance}
\matC_{\vecx} = \alpha^2\lefto(\matC_{\vecz} + \matC_{\vecz\vece} + \matC_{\vecz\vece}^H + \matC_{\vece}\right)
\end{IEEEeqnarray}
where $\matC_\vece = \Ex{}{\vece\vece^H}\in \opC^{BN \times BN}$ is the covariance of $\vece$ and $\matC_{\vecz\vece} = \Ex{}{\vecz\vece^H} \in \opC^{BN \times BN}$; this last matrix can be expressed~also~as
\begin{IEEEeqnarray}{rCl}
\matC_{\vecz\vece} 	
&=& \Ex{}{\vecz\lefto( \alpha^{-1}\vecx - \vecz \right)^H} \\
&=& \Ex{}{\vecz\lefto( \alpha^{-1}\matG\vecz + \alpha^{-1}\vecd - \vecz \right)^H} \\
&=& \matC_{\vecz}(\alpha^{-1}\matG - \matI_{BN}). \label{eq:crosscovar}
\end{IEEEeqnarray}
Here, we have used that $\Ex{}{\vecz\vecd^H} = \matzero_{BN \times BN}$. Inserting~\eqref{eq:crosscovar} into~\eqref{eq:outcovariance}, we obtain
\begin{IEEEeqnarray}{rCl} \label{eq:uncorrcovariance_final}
\matC_{\vecx} = \alpha\lefto(\matG\matC_{\vecz} + \matC_{\vecz}\matG\right) + \alpha^2\lefto(\matC_{\vece} - \matC_{\vecz} \right).
\end{IEEEeqnarray}
Note that the only unknown quantity in \eqref{eq:uncorrcovariance_final} is $\matC_{\vece}$. Not surprisingly, evaluating $\matC_\vece$ is just as difficult as evaluating $\matC_\vecx$ and, in general, no closed-form expression is known. However, if the step size~$\Delta$ of the DACs is set such that the overload distortion is negligible compared to the granular distortion, the error $\vece$ can be accurately approximated by
\begin{IEEEeqnarray}{rCl} 
\vece &=& \alpha^{-1}\quantize(\vecz) - \vecz \approx \rounding(\vecz) - \vecz \label{eq:quant_error_round_approx}
\end{IEEEeqnarray}
where the \emph{rounding function} $\rounding(\cdot)$ is defined as follows:  
\begin{IEEEeqnarray}{rCl} \label{eq:rounding}
\rounding(z) &=&
\begin{cases}
	\Delta \lefto\lfloor \dfrac{z}{\Delta} + \dfrac{1}{2} \right\rfloor, & \text{if } L \text{ is odd}, \\[9pt]
	\Delta \lefto\lfloor \dfrac{z}{\Delta}\right\rfloor + \dfrac{\Delta}{2}, & \text{if } L \text{ is even.}
\end{cases}
\end{IEEEeqnarray}
Some comments on \eqref{eq:quant_error_round_approx} and \eqref{eq:rounding} are in order.
The rounding function $\rounding(\cdot)$ describes a uniform symmetric quantizer with the same step size $\Delta$ as the uniform symmetric quantizer described by $\alpha^{-1}\quantize(\cdot)$ but with an infinite number of quantization levels. 
If the input $z$ lies within the granular region of the uniform quantizer described by $\alpha^{-1}\quantize(\cdot)$, i.e., if $\abs{z} \le A_\text{clip}$, then $\rounding(z) = \alpha^{-1}\quantize(z)$. 
If, however, the amplitude of the input $z$ exceeds the clipping level $A_\text{clip}$, i.e., if $\abs{z} > A_\text{clip}$, then $\rounding(z) \neq \alpha^{-1}\quantize(z)$.
The difference between $\alpha^{-1}\quantize(\cdot)$ and $\rounding(\cdot)$ is illustrated in~\fref{fig:quantizer_and_rounder}.
Replacing~$\alpha^{-1}\quantize(\cdot)$ with $\rounding(\cdot)$ is convenient because the statistical theory of the quantizer~\eqref{eq:rounding} is well-investigated (see, e.g., \cite{widrow96a, sripad77a, lipshitz92a, widrow08a}) and the corresponding covariance matrix~$\matC_\vece$ is known. 
Specifically, let $e_m$ and $e_n$ denote the $m$th and the $n$th entry of $\vece$, respectively.
Furthermore, assume that \eqref{eq:quant_error_round_approx} holds with equality. For a~midtread quantizer ($L$ odd), Sripad and Snyder showed~that~\cite[Eq.~(23)]{sripad77a}
\begin{IEEEeqnarray}{rCl} 
\IEEEeqnarraymulticol{3}{l}{
\Ex{}{e_{m} e_{n}^*} 
} \nonumber\\
&=&  \frac{2\Delta^2}{\pi^2} \sum_{a=1}^\infty \sum_{b = 1}^\infty \frac{(-1)^{a +b }}{ab} \exp\lefto( -\frac{\pi^2\lefto( a^2\sigma_m^2 + b^2\sigma_{n}^2\right)}{\Delta^2} \right) \nonumber \\
&& \times \lefto(\sinh\lefto( \frac{2\pi^2 a b \, \sigma_{m,n}^R}{\Delta^2}\right) + j \sinh\lefto( \frac{2\pi^2 a b \, \sigma_{m,n}^I}{\Delta^2}\right)\right). \IEEEeqnarraynumspace \label{eq:Cqq_nondiag_approx_odd}
\end{IEEEeqnarray}
For midrise quantizers ($L$ even) it further holds that~(see~\fref{app:rounding_nondiag})
\begin{IEEEeqnarray}{rCl} 
\IEEEeqnarraymulticol{3}{l}{
\Ex{}{e_{m} e_{n}^*}
} \nonumber\\
&=& \frac{2\Delta^2}{\pi^2} \sum_{a=1}^\infty \sum_{b = 1}^\infty \frac{1}{a b} \exp\lefto( -\frac{\pi^2\lefto( a^2\sigma_m^2 + b^2\sigma_{n}^2\right)}{\Delta^2} \right) \nonumber \IEEEeqnarraynumspace\\
&& \times \lefto(\sinh\lefto( \frac{2\pi^2 ab \, \sigma_{m,n}^R}{\Delta^2}\right) + j \sinh\lefto( \frac{2\pi^2 ab \,\sigma_{m,n}^I}{\Delta^2}\right)\right). \IEEEeqnarraynumspace \label{eq:Cqq_nondiag_approx_even}
\end{IEEEeqnarray}
Using \eqref{eq:Cqq_nondiag_approx_odd} and \eqref{eq:Cqq_nondiag_approx_even}, we can write the covariance of $\vece$ for every $L \ge 2$ as
\begin{IEEEeqnarray}{rCl} \label{eq:Cqq_rounding}
	\matC_\vece &=& \frac{2\Delta^2}{\pi^2} \sum_{a=1}^\infty \sum_{b = 1}^\infty  \frac{\cos(\pi L)^{a + b}}{a b} \nonumber\\
&& \times \exp\lefto( -\frac{\pi^2 a^2}{\Delta^2}  \text{diag}\lefto(\matC_\vecz\right)\mathbf{1}_{BN \times BN}\right. \nonumber\\
&& \lefto. -\frac{\pi^2 b^2}{\Delta^2}\mathbf{1}_{BN \times BN}\,\text{diag}\lefto(\matC_\vecz\right) \right) \nonumber\\
&& \odot  \lefto(\sinh\lefto( \frac{2\pi^2 a b }{\Delta^2}\,\Re\{ \matC_\vecz\}\right) \right. \nonumber\\
&& \lefto. + j \sinh\lefto( \frac{2\pi^2 a b}{\Delta^2}\,\Im\{ \matC_\vecz\}\right)\right).
\end{IEEEeqnarray}
We obtain the desired rounding approximation for $\matC_\vecx$ by inserting~\eqref{eq:Cqq_rounding} into \eqref{eq:uncorrcovariance_final}.
Interestingly, for the case $L=2$, it is possible to retrieve Van Vleck's arcsine law~\eqref{eq:arcsine_mimo} from~\eqref{eq:uncorrcovariance_final} and~\eqref{eq:Cqq_rounding}. We formalize this result in the following theorem; a proof is given in~\fref{app:proof_conv}.
\begin{thm} \label{thm:conv}
Assume $L=2$, insert~\eqref{eq:Cqq_rounding} into \eqref{eq:uncorrcovariance_final}, and let $\Delta \rightarrow \infty$. Then, we obtain Van Vlecks's arcsine law~\eqref{eq:arcsine_mimo}.
\end{thm}

Finally, by inserting~\eqref{eq:Cqq_rounding} and~\eqref{eq:uncorrcovariance_final} into \eqref{eq:Cdd}, we obtain the rounding approximation $\matC_\vecd^\text{round}$ of $\matC_\vecd$:
\begin{IEEEeqnarray}{rCl} \label{eq:Cdd_rounding}
	\matC_\vecd^\text{round} &=& \frac{2\alpha^2\Delta^2}{\pi^2} \sum_{a=1}^\infty \sum_{b = 1}^\infty  \frac{\cos(\pi L)^{a + b}}{a b} \nonumber\\
&& \times \exp\lefto( -\frac{\pi^2 a^2}{\Delta^2}  \text{diag}\lefto(\matC_\vecz\right)\mathbf{1}_{BN \times BN}\right. \nonumber\\
&& \lefto. -\frac{\pi^2 b^2}{\Delta^2}\mathbf{1}_{BN \times BN}\,\text{diag}\lefto(\matC_\vecz\right) \right) \nonumber\\
&& \odot  \lefto(\sinh\lefto( \frac{2\pi^2 a b }{\Delta^2}\,\Re\{ \matC_\vecz\}\right) \right. \nonumber\\
&& \lefto. + j \sinh\lefto( \frac{2\pi^2 a b}{\Delta^2}\,\Im\{ \matC_\vecz\}\right)\right) \nonumber\\[5pt]
&& + \alpha\lefto( \matG\matC_\vecz + \matC_\vecz\matG\right) - \alpha^2\matC_\vecz - \matG\matC_\vecz\matG .\IEEEeqnarraynumspace
\end{IEEEeqnarray}
As we will see in~\fref{sec:numerical}, the rounding approximation~\eqref{eq:Cdd_rounding} is accurate independently of the resolution of the DACs, provided that $\Delta$ is chosen so that clipping occurs with low probability. 

Note that evaluating~\eqref{eq:Cdd_rounding} involves computing two infinite sums. 
The series, however, converges rapidly because the terms decay exponentially in the variables $a$ and $b$.
For the numerical results reported in \fref{sec:numerical}, we evaluate~\eqref{eq:Cdd_rounding} by summing only over the first $30$ terms of both sums; this turns out to be sufficient to obtain accurate~results.

\subsection{Diagonal Approximation}
\label{sec:quantizer_lmmse}

Recall from~\fref{sec:quantizer_exact} that the diagonal elements of $\matC_\vecx$ can be computed exactly using~\eqref{eq:xm2}. 
Building on this observation, we present next a \emph{diagonal approximation} $\matC_\vecd^\text{diag}$ of $\matC_\vecd$ in which the distortion caused by the DACs is modeled as a white process, both in space and time, by assuming that the off-diagonal elements of $\matC_\vecd$ are zero.~Specifically,
\begin{IEEEeqnarray}{rCl} 
\matC_{\vecd}^\text{diag} 
&=& \text{diag}\lefto( \matC_\vecx \right) - \matG \, \text{diag}\lefto(\matC_\vecz \right)\matG \label{eq:Cdd_approx_LMMSE_temp}\\[5pt]
&=& \frac{\alpha^2\Delta^2}{2}\lefto(L-1\right)^2\matI_{BN} - \matG\,\text{diag}\lefto(\matC_\vecz \right)\matG \nonumber\\ 
&& - 4\alpha^2\Delta^2\sum_{i=1}^{L-1}\lefto(i-\frac{L}{2}\right) \nonumber\\
&& \times\Phi\lefto(\sqrt{2}\,\text{diag}\lefto(\matC_\vecz\right)^{-\frac{1}{2}}\lefto(i-\frac{L}{2}\right)\right). \label{eq:Cdd_approx_LMMSE}
\end{IEEEeqnarray}
Here, we used~\eqref{eq:Cdd} to obtain \eqref{eq:Cdd_approx_LMMSE_temp}, and~\eqref{eq:xm2} to obtain~\eqref{eq:Cdd_approx_LMMSE}. 
Note that $\matC_\vecd^\text{diag} = \matC_\vecd \odot \matI_{BN}$ (cf.~\cite[Eq.~(26)]{bjornson18a}), which implies that the elements on the diagonal of $\matC_\vecd$ in~\eqref{eq:Cdd_approx_LMMSE}, for Gaussian inputs, are~exact. 

For the uplink case, an approximate model called the \emph{additive quantization noise model}~(AQNM) is commonly used to characterize the quantization distortion caused by low-resolution ADCs (see, e.g.,~\cite{orhan15a, fan15a}). 
Similarly to the diagonal approximation in~\eqref{eq:Cdd_approx_LMMSE}, in the AQNM the covariance matrix of the quantization distortion is approximated by a diagonal matrix. 
The AQNM model, however, assumes that the the power of the input to the quantizer is equal on all antennas (which can be achieved in the uplink by using an automatic gain control circuit) and that quantization labels are rescaled to minimize the MSE between the quantizer input and output.
These assumptions are not valid for the massive MU-MIMO-OFDM downlink scenario considered is this paper. Indeed, the power of the transmitted signal on different antennas is not equal and the quantization labels are uniformly separated.\footnote{For the uplink case, Mezghani and Nossek derived an approximation for the covariance matrix of the quantization distortion caused by symbol-rate-sampling ADCs that, similarly to the proposed rounding approximation, takes into account the correlation in the signal transmitted on different antennas~\cite{mezghani12b}. The approximation in~\cite{mezghani12b}, however, assumes that the power of the input to the quantizer is equal on all antennas and that the quantization labels are rescaled to minimize the MSE between the quantizer input and output (these are the same assumptions that are used to derive the AQNM approximation~\cite{orhan15a, fan15a}). Neither of these assumptions are valid for the massive MU-MIMO-OFDM downlink scenario considered here.}

Despite its simplicity, the diagonal approximation~\eqref{eq:Cdd_approx_LMMSE} is accurate for DACs with medium-to-high resolution and if the OSR is relatively small (e.g., when $L \ge 4$ and~$\osr \le 4$). 
However, as we will demonstrate in~\fref{sec:numerical}, the diagonal approximation is not sufficiently accurate for DACs with low resolution and if the OSR is high (e.g., when $L < 4$ and $\osr > 4$), which implies that the correlation within the distortion caused by the DACs should not be ignored.

%

%
%
%
%
%
%

\section{Numerical Results}
\label{sec:numerical}

We focus on a massive MU-MIMO-OFDM system in which the number of BS antennas is $B = 128$ and the number of UEs is $U = 16$.\footnote{Our simulation framework is available for download from GitHub (\url{https://github.com/quantizedmassivemimo/1bit_linear_precoding_ofdm}). The purpose is to enable~interested readers to perform simulations with different system~parameters than the ones reported here.} We consider a frequency-selective Rayleigh fading channel with $T = 4$ nonzero taps and a uniform power delay profile. 
The OFDM parameters are inspired by a $5$~MHz LTE system~\cite{3gpp17a}. Specifically, the number of occupied subcarriers is $S = 300$ and the subcarrier spacing is $\Delta f = 15$~kHz. Furthermore, we assume that the occupied subcarriers are the first $150$ to the left and to the right of the DC subcarrier (the DC subcarrier is not occupied).
The total number of subcarriers (the size of the~DFT) is $N = 1024$. Hence, the sampling rate of the DACs is $f_s = N\Delta f = 1024 \cdot 15 \cdot 10^3 = 15.36$~MHz and the OSR is~$\osr=N/S=1024/300\approx3.4$.\footnote{In~\fref{sec:osr}, we shall investigate the impact of the OSR on performance.}

Under the assumption that the input to the DACs on the $b$th antenna ($b = 1, 2, \dots, B$) is~$z_b \sim \jpg\lefto(0, P/(\osr B) \right)$ and that the clipping level is set to $A_\text{clip} = \sqrt{P/(2\osr B)}\,\lefto(1 - \Phi^{-1}\lefto(P_\text{clip}/2\right)\right)$, the DACs will clip the signal with probability $P_\text{clip}$. The corresponding step size is $\Delta = 2A_\text{clip}/L$. In what follows, we have set the clipping level of the DACs so that $P_\text{clip} = 0.1\%$. This choice is not necessarily optimal, but ensures that the clipping distortion caused by the DACs is small compared to the granular distortion for the values of~$L$ considered here, which will enable us to use the rounding approximation~\eqref{eq:Cqq_rounding}. Furthermore, we shall see that this particular choice yields near-optimal (infinite resolution) performance also for DACs with low~resolution.

\begin{figure*}
\centering
\subfloat[Transmitted, 1-bit DACs ($L=2$).]{\includegraphics[width=.3\textwidth]{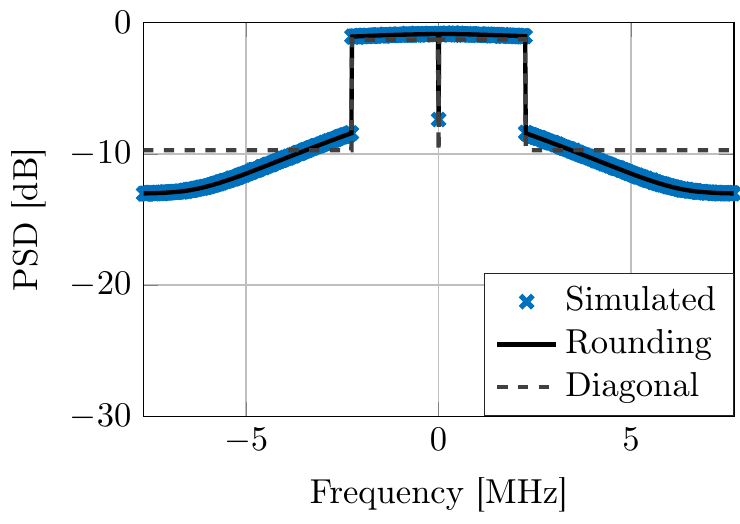}\label{fig:psd_1bit_tx}} \quad
\subfloat[Transmitted, 2-bit DACs ($L=4$).]{\includegraphics[width=.3\textwidth]{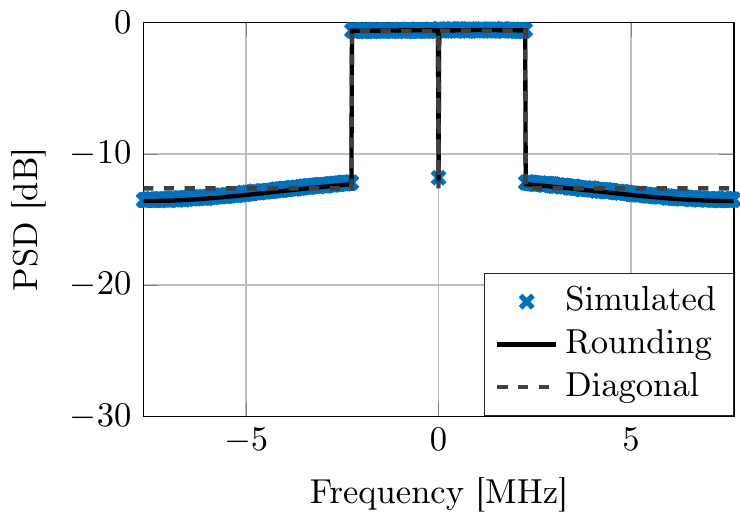}} \quad
\subfloat[Transmitted, 3-bit DACs ($L=8$).]{\includegraphics[width=.3\textwidth]{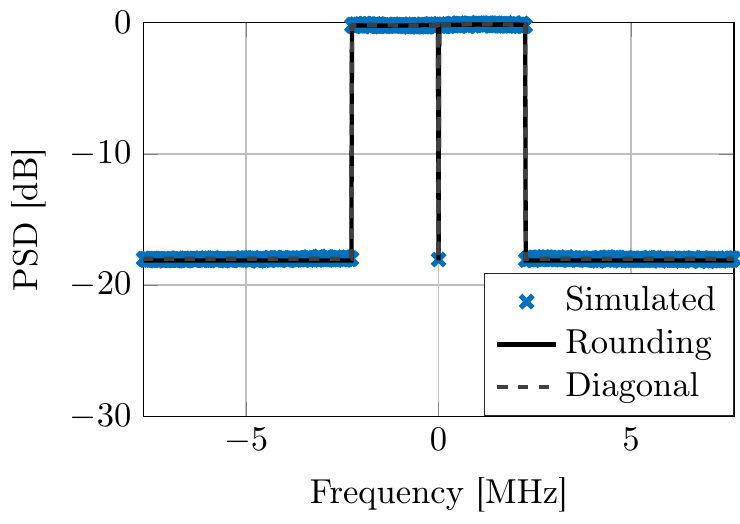}\label{fig:psd_3bit_tx}} \\
\subfloat[Received, 1-bit DACs ($L=2$).]{\includegraphics[width=.3\textwidth]{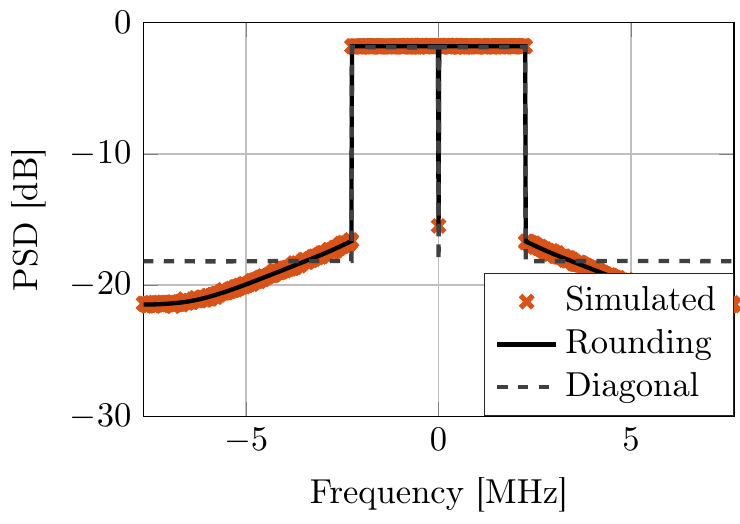}} \quad
\subfloat[Received, 2-bit DACs ($L=4$).]{\includegraphics[width=.3\textwidth]{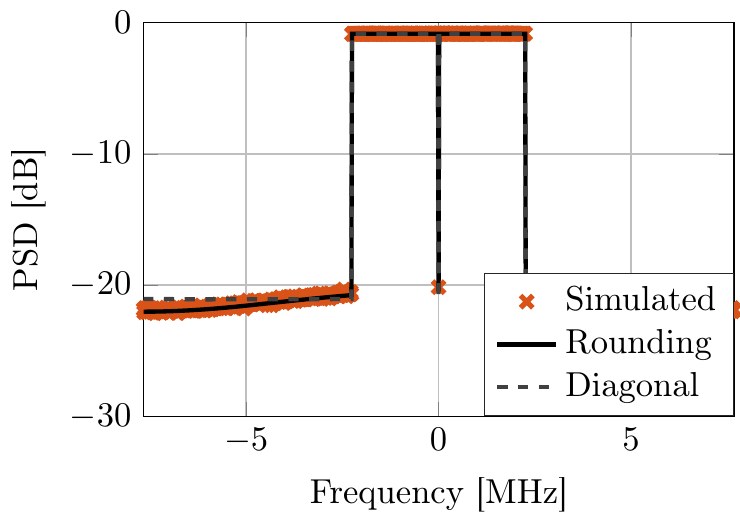}} \quad
\subfloat[Received, 3-bit DACs ($L=8$).]{\includegraphics[width=.3\textwidth]{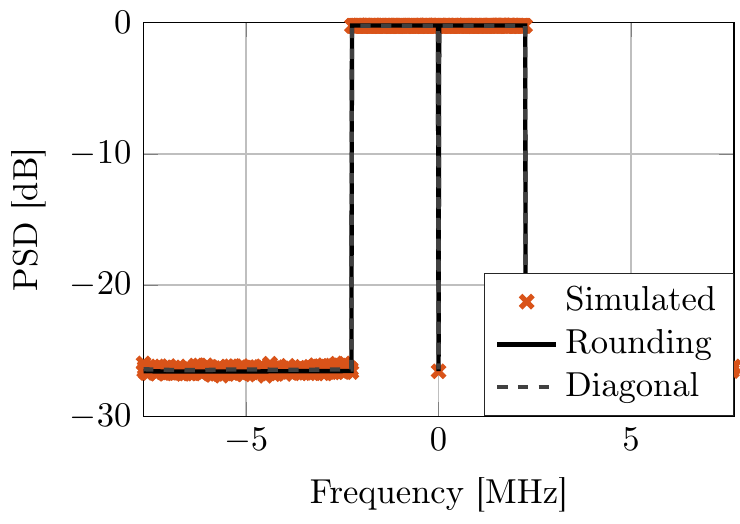}\label{fig:psd_3bit_rx}} 
\caption{PSD of the transmitted and received signal; $B = 128$ and $U=16$. The markers correspond to simulated values, the solid lines correspond to the rounding approximation presented in~\fref{sec:quantizer_rounding} and the dashed lines correspond to the diagonal approximation presented in~\fref{sec:quantizer_lmmse}.}
\label{fig:psd}
\end{figure*}

\subsection{Power Spectral Density} \label{sec:psd}

To demonstrate the accuracy of the approximations in \fref{sec:quantizer_rounding} and \fref{sec:quantizer_lmmse}, we plot in \fref{fig:psd} the (normalized) PSD of the transmitted signal (averaged over the BS antennas and over $100$ channel realizations) and the (normalized) PSD of the received signal (averaged over the UEs and over $100$ channel realizations).
Here, the BS uses ZF precoding and the data vectors~$\vecs_k$ for $k \in \setS_d$ contain QPSK~symbols.\footnote{Recall that in~\fref{sec:performance} and~\fref{sec:quantization}, we assumed that the per-antenna DAC input is a zero-mean Gaussian random variable. This assumption holds approximately true also for the case of finite-cardinality constellations and linear precoding, because the per-antenna DAC input can be written as sum of $US$ independent and identically distributed random variables with zero mean and finite variance. Hence, as $US$ grow large, the per-antenna DAC input converges to a Gaussian random variable by the central limit theorem.}
Numerical simulations are compared with analytic results obtained by computing the PSD of the transmitted vector $\vecx$ and of the corresponding received vector~$\vecy$ using the rounding approximation and the diagonal approximation presented in~\fref{sec:quantizer_rounding} and \fref{sec:quantizer_lmmse}, respectively. 
We note that the rounding approximation yields accurate results independently of the resolution of the DACs. We also note that the diagonal approximation results in a poor approximation of the PSD for the case $L=2$ but yields a more accurate approximation as the number of bits increase. 
Indeed, when the number of bits increases, the distortion caused by the DACs becomes more spectrally white (compare \fref{fig:psd_1bit_tx} and \fref{fig:psd_3bit_tx}).
We also see from the figure that the low-resolution DACs cause severe OOB distortion at the~BS, which is captured accurately by the rounding approximation for all values of $L$. 
Interestingly, the relative amount of OOB distortion is smaller at the UEs than at the BS, which is in line with recent findings reported for PAs in~\cite{mollen17a}.
Nevertheless, the OOB distortion caused by the low-resolution DACs is a significant issue in practical systems as it may cause interference to UEs operating in adjacent frequency~bands. 

\subsection{Error-Rate Performance}


\subsubsection{Uncoded BER}

\begin{figure}[t!]
\centering
\subfloat[Uncoded BER with QPSK; $B = 1$ and $U=1$.]{\includegraphics[width = \figwidth]{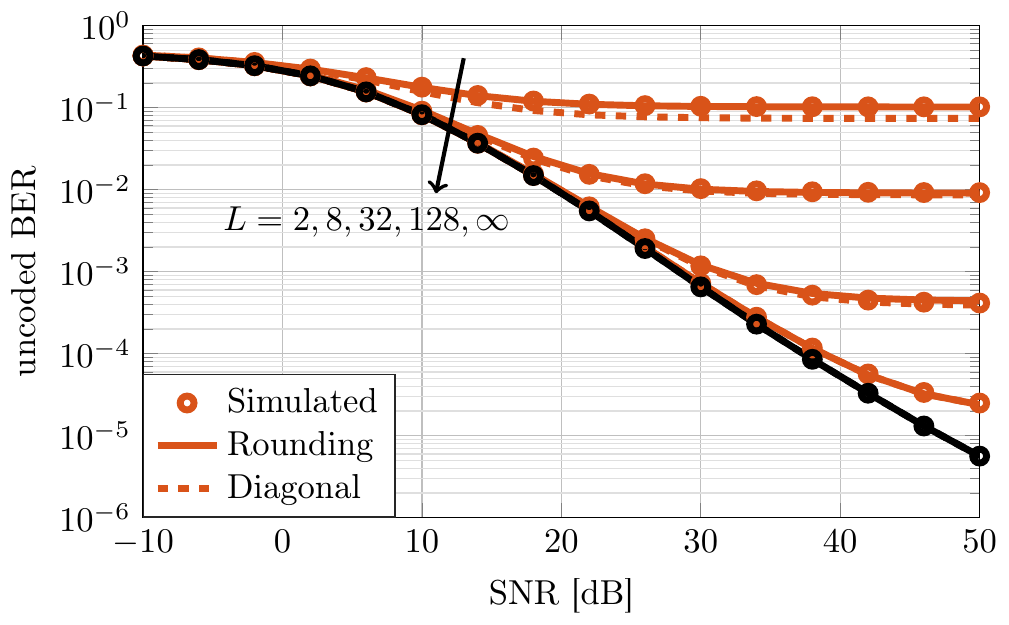}
\label{fig:siso_ber_uncoded}} \quad
\subfloat[Uncoded BER with QPSK; $B = 128$ and $U=16$.]{\includegraphics[width = \figwidth]{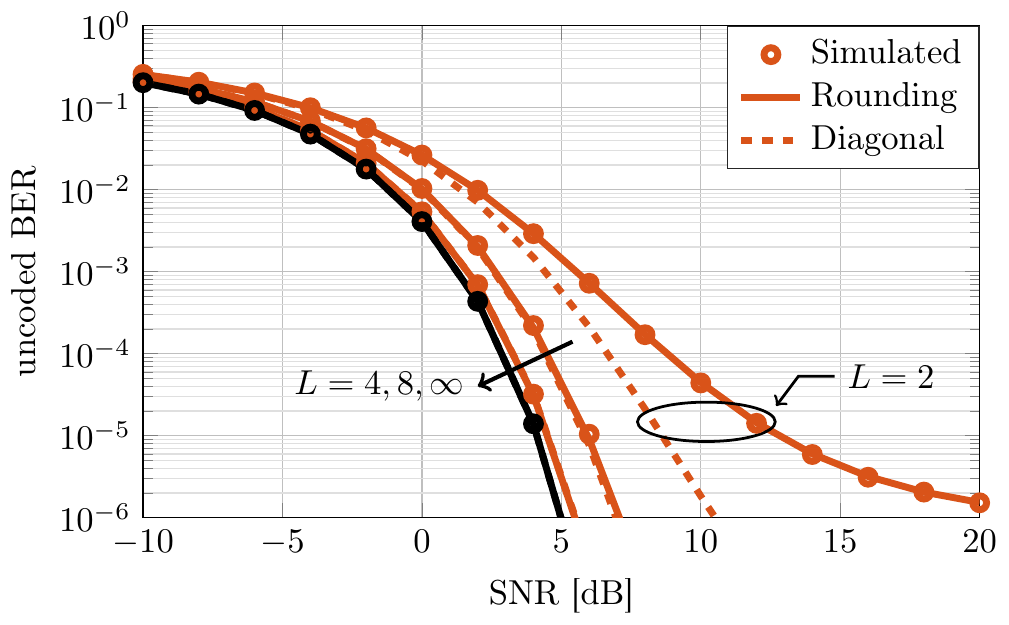}\label{fig:mimo_ber_uncoded}}
\caption{Uncoded BER with ZF and QPSK signaling. The markers correspond to simulated values, the solid lines correspond to the BER~\eqref{eq:ber_qpsk} computed using the rounding approximation in~\fref{sec:quantizer_rounding}, and the dashed lines correspond to the BER~\eqref{eq:ber_qpsk} computed using the diagonal approximation in~\fref{sec:quantizer_lmmse}. The black lines correspond to the infinite-resolution~case.}
\label{fig:siso_ber}
\end{figure}

If the elements of the symbol vector $\vecs_k$ for $k \in \setS_d$ are drawn independently from a QPSK constellation, we can approximate the uncoded BER by
\begin{IEEEeqnarray}{rCl} \label{eq:ber_qpsk}
	\textit{BER} = 1 - \frac{1}{US} \sum_{k \in \setS_d} \sum_{u=1}^U  \Ex{}{  \Phi\lefto(\sindr_{u,k}^{1/2}(\widehat\matH)\right)} 
\end{IEEEeqnarray}
where~$\sindr_{u,k}(\widehat\matH)$ is given in \eqref{eq:sindr_mimo}. We evaluate this quantity by using the rounding approximation~\eqref{eq:Cdd_rounding} and the diagonal approximation~\eqref{eq:Cdd_approx_LMMSE}.
Note that for the approximations to be accurate, the number of BS antennas~$B$ does not need to be large. To illustrate this aspect, we show in~\fref{fig:siso_ber_uncoded} the uncoded BER with QPSK and ZF for the single-input single-output (SISO)~case (i.e, when $U = 1$ and $B=1$) as a function of the SNR and the number of DAC bits.\footnote{In the SISO-OFDM case, ZF precoding reduces to channel inversion.} 
We compare simulated BER values with the analytical BER in~\eqref{eq:ber_qpsk} and note that the rounding approximation is accurate over the entire range of SNR values. We note that the diagonal approximation become more accurate as the number of DAC bits increase.
We further note that, in the SISO-OFDM case, low uncoded BERs are not supported with QPSK and low-resolution DACs. Indeed, $7$--$8$ DAC bits are required to achieve a target BER of $10^{-4}$ without a significant performance degradation compared to the infinite-resolution case.

\begin{figure}[t]
\centering
\subfloat[Coded BER with QPSK; $B = 128$ and $U=16$.]{\includegraphics[width = \figwidth]{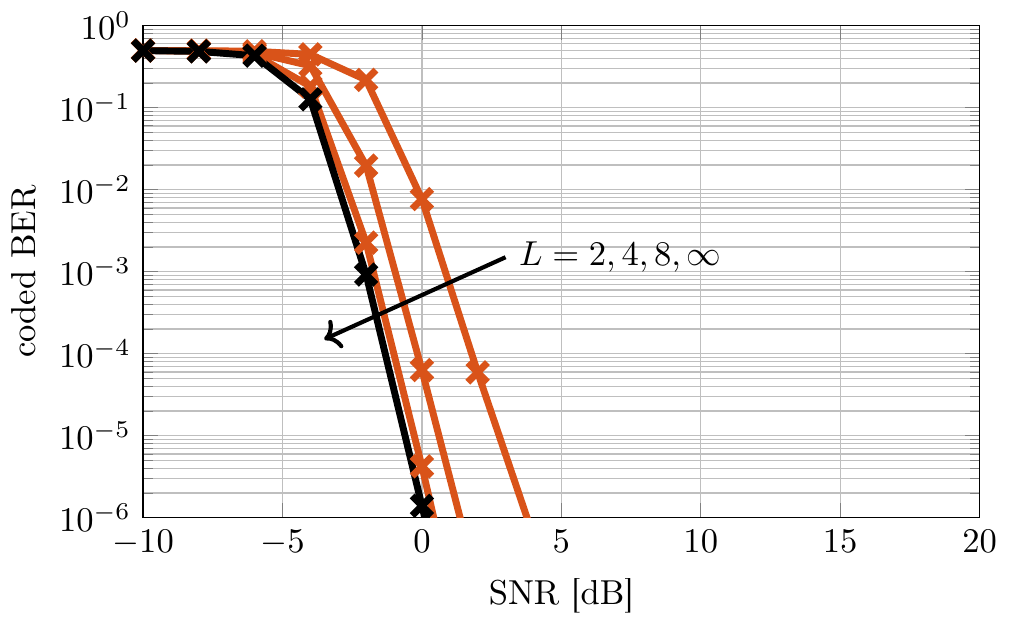}\label{fig:mimo_ber_qpsk_coded}} \quad
\subfloat[Coded BER with 16-QAM; $B = 128$ and $U=16$.]{\includegraphics[width = \figwidth]{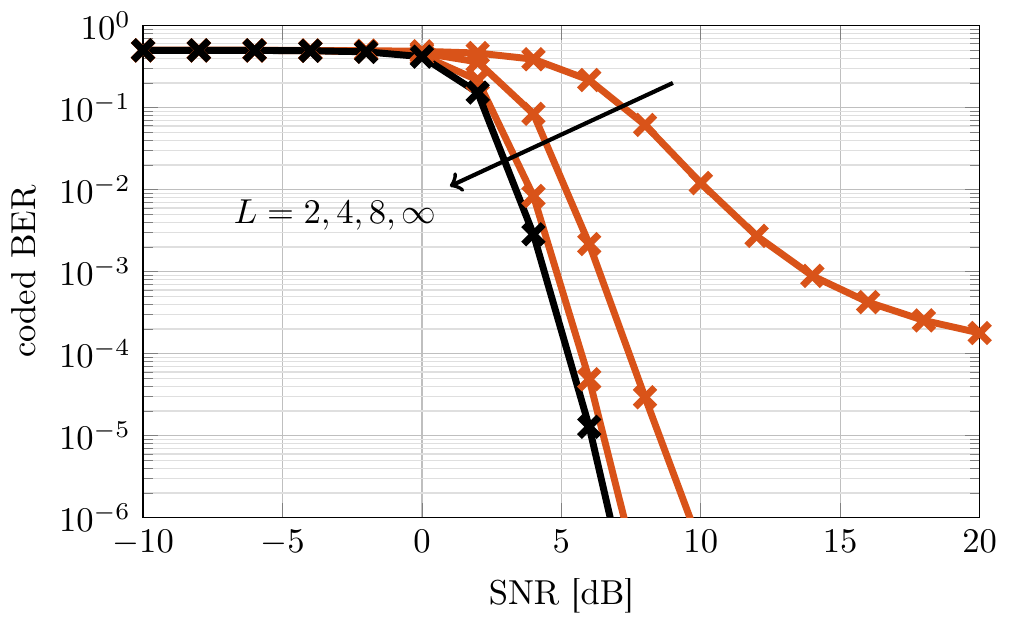}\label{fig:mimo_ber_16qam_coded}}
\caption{Coded BER (simulated) with ZF for the case of QPSK and 16-QAM signaling. The information streams are encoded using a rate-5/6 convolutional code spanning 10 OFDM symbols. The UEs use nearest-neighbor detection and soft-input max-log BCJR decoding. The black lines correspond to the infinite-resolution case.}
\label{fig:mimo_coded_ber}
\end{figure}

In the massive MU-MIMO-OFDM case, the large number of antennas at the BS enables a considerable reduction of the resolution of the DACs compared to the SISO-OFDM~case. 
To illustrate this, we show in \fref{fig:mimo_ber_uncoded} the uncoded BER with QPSK and ZF as a function of the SNR and the number of DAC bits for the case $B = 128$ and $U = 16$. 
In contrast to the SISO-OFDM case, low uncoded BERs are now supported by low-resolution DACs. Indeed, an uncoded BER below $10^{-4}$ is supported in the 1-bit-DAC case provided that the SNR exceeds $9$~dB. Furthermore, only $3$--$4$ DAC bits are necessary to approach  infinite-resolution performance for a target BER of~$10^{-4}$.  
We again note that the rounding approximation~\eqref{eq:Cdd_rounding} is accurate over the entire range of SNR values and independently of the number of DAC bits, and that the diagonal approximation~\eqref{eq:Cdd_approx_LMMSE} becomes more accurate as the number of DAC bits~increase. 
However, the diagonal approximation significantly overestimates the BER performance with 1-bit DACs for high values of SNR, as this approximation does not take into account the inherent (spatial and temporal) correlation in the DAC distortion. 
Hence, one must be careful when using approximations that ignore the correlation in the quantization distortion as they are accurate only in some scenarios (e.g., for low SNR or when the resolution of the quantizer is sufficiently high).  
%

\subsubsection{Coded BER}

In \fref{fig:mimo_coded_ber}, we show the coded BER with ZF as a function of the SNR and the number of DAC bits for the case of QPSK and 16 quadrature amplitude modulation~(QAM) signaling. Here, we consider the case $B=128$ and $U = 16$, and show only simulated BER values. The BS uses a (weak) rate-5/6 convolutional code with random interleaving to encode the information bits (separately for each UE) over $10$ OFDM symbols. Hence, a codeword spans $3000$~symbols. Each UE performs soft-input max-log Bahl, Cocke, Jelinek and Raviv~(BCJR) decoding to estimate the transmitted information~streams. 
We note that high-order constellations, such as 16-QAM, are supported with linear precoding in the massive MU-MIMO-OFDM case despite the low-resolution DACs, and that only few DAC bits are needed to close the gap to the infinite-resolution~performance. 

In this work, for analytical tractability, we have focused exclusively on linear precoding. However, it is well-known (see, e.g.,~\cite{jacobsson17d, jacobsson16d, jedda16a, swindlehurst17a, castaneda17a, landau17a, jacobsson18b, li18a, shao18a, shao18b, sohrabi18a}) that is possible to achieve superior performance, at the cost of an increased computational complexity, by using nonlinear precoders. For the case of OFDM transmission with 1-bit DACs, a comparison between linear and nonlinear precoding, in terms of uncoded/coded BER and computational complexity, is provided in~\cite{jacobsson17f}.

\subsection{Achievable Rate}

\begin{figure}[t]
\centering
	\includegraphics[width = \figwidth]{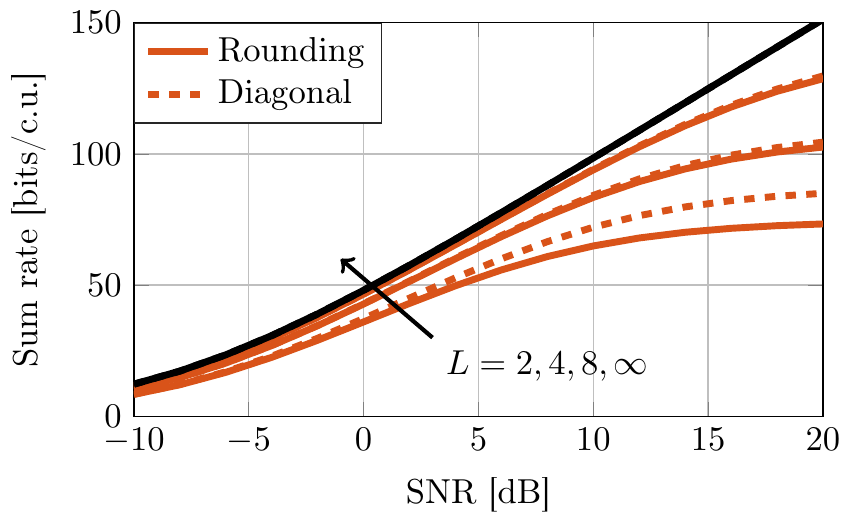}
	\caption{Sum-rate with ZF and Gaussian signaling; $B = 128$ and $U = 16$. The solid lines correspond to the rounding approximation in~\fref{sec:quantizer_rounding}, the dashed lines correspond to the diagonal approximation in~\fref{sec:quantizer_lmmse}. The black lines, which overlap, correspond to the sum rate achievable in the infinite-resolution~case.}	
	\label{fig:rate}
\end{figure}


In \fref{fig:rate}, we show the achievable sum rate with Gaussian signaling and ZF precoding as a function of the SNR and the number of DAC levels. 
Recall that this sum rate is achieved by using a Gaussian codebook and nearest-neighbor decoding at the UEs (see~\fref{sec:achievable}).
The sum rate in~\eqref{eq:rate} is evaluated by computing \eqref{eq:sindr_mimo} using the rounding approximation and the diagonal approximation in \eqref{eq:Cdd_rounding} and \eqref{eq:Cdd_approx_LMMSE}, respectively. 
%
%
%
We note that the diagonal approximation is accurate for the case $L \ge 4$ but significantly overestimates  the achievable rate for $L=2$. We further note that high sum-rate throughputs are supported by the massive MU-MIMO-OFDM system, despite the low-resolution DACs at the~BS.

\subsection{Impact of Imperfect CSI}
\label{sec:csi_error}

\begin{figure}[t]
\centering
	\subfloat[Impact of Imperfect CSI; $\snr = 5$~dB, $\xi = 3.4$.]{\includegraphics[width = \figwidth]{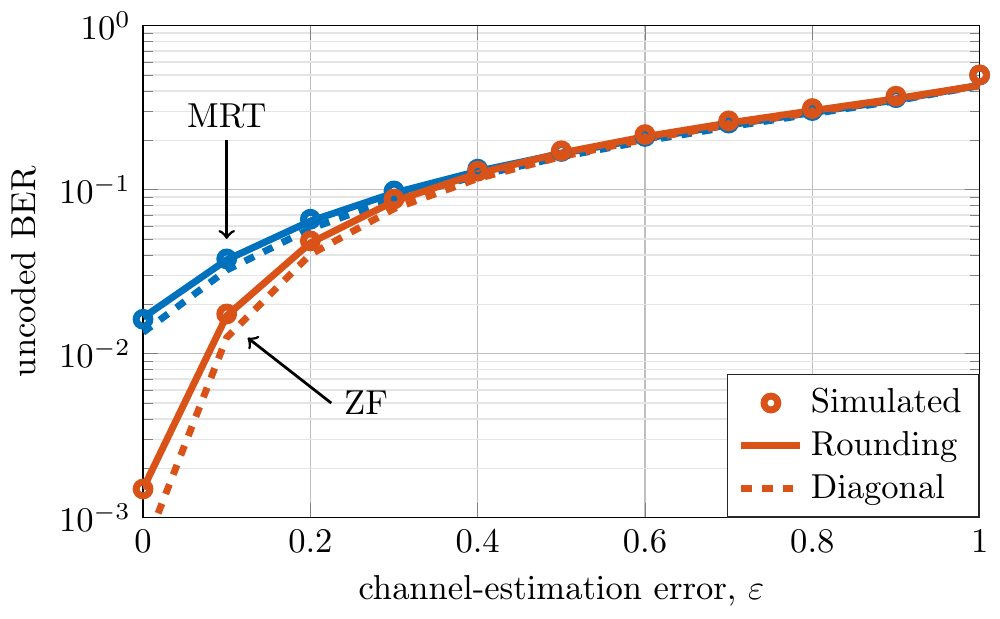}\label{fig:csi_error}} \quad
	\subfloat[Impact of OSR; $\snr = 10$~dB, $\varepsilon = 0$.]{\includegraphics[width = \figwidth]{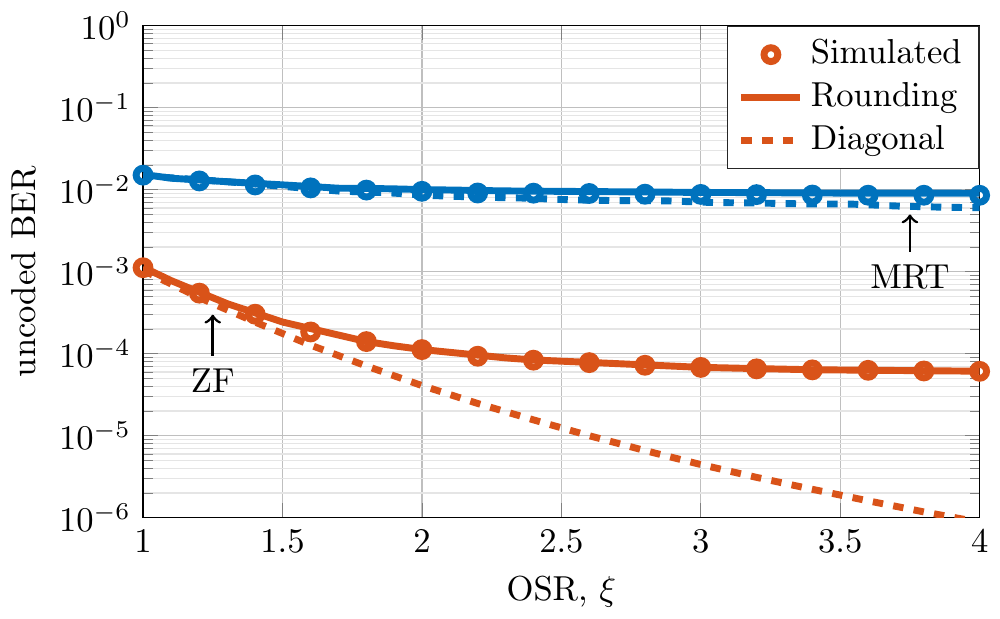}\label{fig:osr}}
	\caption{Impact of the imperfect CSI and OSR on the uncoded BER for the 1-bit-DAC case; $B = 128$ and $U = 16$. The markers correspond to simulated values, the solid lines correspond to the BER~\eqref{eq:ber_qpsk} according to the rounding approximation in~\fref{sec:quantizer_rounding}, the dashed lines correspond to the BER~\eqref{eq:ber_qpsk} according to the diagonal approximation in~\fref{sec:quantizer_lmmse}.}	
\end{figure}

Until now, we have assumed that perfect CSI is available at the BS. We now relax this assumption and investigate the impact on performance of imperfect CSI.
Specifically, we consider the case in which the BS has access only to noisy versions $\{ \matH_t^\text{est} \}$ of~$\{ \matH_t \}$ for $t = 0, 1, \dots, T-1$. Specifically, $\matH_t^\text{est} = \sqrt{1-\varepsilon}\matH_t + \sqrt{\varepsilon}\matH_t^\text{err}$ for $t = 0, 1, \dots, T-1$, where $  \varepsilon  \in [0,1]$ and where the entries of $\matH_t^\text{err}$ are uncorrelated with zero mean and unit variance.
The corresponding frequency-domain estimate for the $k$th subcarrier ($k \in \setS_d$) is $\widehat\matH_{k}^\text{est} = \sum_{t=0}^{T-1} \matH_t^\text{est} \exp\lefto( - jk\frac{2\pi}{N}t \right)$; the corresponding MRT and ZF precoding vectors are obtained from this estimate using~\eqref{eq:Pmrt} and~\eqref{eq:Pzf}, respectively.
 The case $\varepsilon = 0$ corresponds to perfect CSI (i.e., the case considered in~\fref{fig:mimo_ber_uncoded}) and the case $\varepsilon = 1$ corresponds to no CSI.
%
In a time-division duplex system, the value of the channel-estimation error $\varepsilon$ will depend on the pilot sequence that have been transmitted during the uplink phase and on the resolution of the ADCs at the BS~(see, e.g.,~\cite{studer16a,mollen16c, jacobsson17b, li17b}). 
In what follows, for simplicity, we assume that the entries of $\matH_t^\text{err}$ are $\jpg(0,1)$ distributed for~$t = 0,1,\dots, T-1$.
In~\fref{fig:csi_error}, we show, for the 1-bit-DAC case, the uncoded BER with QPSK as a function of the channel-estimation error $\varepsilon$ for MRT and ZF precoding. We note that the rounding approximation~is accurate also for the case of imperfect CSI. We further note that ZF outperforms MRT for~$\varepsilon < 0.4$.

\subsection{Impact of Oversampling}
\label{sec:osr}

In~\fref{fig:osr}, we investigate the impact of the OSR on the uncoded BER for the case of 1-bit DACs and perfect CSI. Specifically, we plot the uncoded BER for the case of uncoded QPSK with MRT and ZF precoding as a function of the OSR. The SNR is set to $\snr = 10$~dB. We note that, for ZF, the uncoded BER can be considerably improved by operating the 1-bit DACs at a sampling rate higher than the symbol rate. Indeed, the uncoded BER with ZF can be decreased by an order of magnitude compared to the symbol-rate sampling case ($\osr = 1$) by operating the 1-bit DACs at twice the symbol rate. However, further increasing the sampling rate yields only marginal performance~gains. 
For MRT, the OSR has little impact on BER performance as, in this case, the system is limited by the MU interference rather than by the DAC distortion.
We also note that, in the 1-bit-DAC case, the diagonal approximation is accurate for small OSRs (e.g., for~$\osr < 1.4$) but underestimates significantly the BER for larger values of~$\osr$.  For higher resolution DACs, the diagonal approximation is accurate also for high OSRs (see, e.g.,~\fref{fig:mimo_ber_uncoded}).

\section{Conclusions}
\label{sec:conclusion}

We have characterized the performance in terms of uncoded/coded BER and achievable sum rate of a massive MU-MIMO-OFDM downlink system, in which the BS is equipped with finite-resolution DACs and uses linear precoding. Using Bussgang's theorem, we have derived a lower bound on the achievable rate and an accurate approximation for the uncoded BER with~QPSK. 


We have developed {two} approximations for the distortion caused by the low-resolution DACs. 
The \emph{rounding approximation} ignores the overload distortion caused by the DACs and is accurate for DACs of arbitrary resolution and for any OSR. We also prove that the rounding approximation can be made exact in the 1-bit-DAC case by taking $\Delta$ in~\eqref{eq:Cqq_rounding} to~infinity.
The \emph{diagonal approximation} is an easy-to-evaluate approximation that assumes that the DAC distortion is spatially and temporally white. This approximation is accurate for moderate-to-high DAC resolutions and if the OSR is not too high, but significantly overestimates the performance (in terms of BER and achievable rate) for low-resolution (e.g., 1-bit) DACs. Our results highlight the importance of taking into account the correlation of the distortion caused by the quantizer in the DACs. Similar findings have been  reported recently in~\cite{li17e} for the distortion caused by ADCs in the uplink.

In practice, OOB emissions caused by the low-resolution DACs could prevent their use in practical systems. In this work, for simplicity, we have assumed that the reconstruction stage in the DACs is an ideal low-pass filter.
In the recent paper~\cite{jacobsson17a}, we have analyzed, using the models developed in this paper, spectral and spatial emissions under more realistic assumptions on the reconstruction stage in the~DACs.

\appendices

\section{Derivation of \eqref{eq:Cqq_nondiag_approx_even}}
\label{app:rounding_nondiag}

It can be shown~(see, e.g.,~\cite[Eq.~(9.7)]{widrow08a}) that for midtread rounding quantizers it holds that
\begin{IEEEeqnarray}{rCl}
\Ex{}{e_m^C e_n^R} &=&  \frac{\Delta^2}{2\pi^2}\sum_{a = 1}^\infty \sum_{b=1}^\infty	 \frac{(-1)^{a + b}}{a b} \nonumber\\ 
&& \times\lefto(\Re\lefto\{ \varphi_{z_m^C, z_n^R}\lefto( \frac{2\pi a}{\Delta}, -\frac{2\pi b}{\Delta}\right)\right\} \right. \nonumber\\
&& - \lefto. \Re\lefto\{ \varphi_{z_m^C, z_n^R}\lefto( \frac{2\pi a}{\Delta}, \frac{2\pi b}{\Delta}\right)\right\} \right) \IEEEeqnarraynumspace
\end{IEEEeqnarray}
where $\varphi_{z_m^C, z_n^R}\lefto( u, v \right) = \opE\big[e^{ j \lefto(u z_m^C + v z_n^R\right)}\big]$ is the characteristic function of $[z_m^C, z_n^R]^T$, for~$C \in \{R, I\}$. 

Note that by adding a constant $\Delta/2$ to the input of a midtread rounding quantizer, the output of said midtread rounding quantizer equals exactly the output of a midrise rounding quantizer for the case when no constant has been added. Further note that $\varphi_{z_m^C + \Delta/2, z_n^R + \Delta/2}\lefto( u, v \right) = e^{j \Delta(u + v)/2}\varphi_{z_m^C, z_n^R}\lefto( u, v \right)$. Hence, for midrise rounding quantizers, it holds that
\begin{IEEEeqnarray}{rCl}
\Ex{}{e_m^C e_n^R} &=&	\frac{\Delta^2}{2\pi^2}\sum_{a = 1}^\infty \sum_{b=1}^\infty \frac{(-1)^{a + b}}{a b}  \nonumber\\
&&\times \lefto(\Re\lefto\{ e^{j\pi\lefto( a - b\right)} \varphi_{z_m^C, z_n^R}\lefto( \frac{2\pi a}{\Delta}, -\frac{2\pi b}{\Delta}\right)\right\} \right. \nonumber\\
&& - \lefto. \Re\lefto\{ e^{j\pi\lefto( a + b\right)}\varphi_{z_m^C, z_n^R}\lefto( \frac{2\pi a}{\Delta}, \frac{2\pi b}{\Delta}\right)\right\} \right). \IEEEeqnarraynumspace \label{eq:error1_CR_midrise}
\end{IEEEeqnarray}
In our case, $\big[z_m^C, z_n^R\big]^T$ is a zero-mean Gaussian random vector for which it holds that $\varphi_{z_m^C, z_n^R}\lefto( u, v \right) = \exp\lefto( -\frac{1}{4}\lefto( u \sigma_m^2 + 4uv\sigma_{m,n}^C +  v \sigma_n^2 \right)\right)$, which we use to simplify~\eqref{eq:error1_CR_midrise} as follows:
\begin{IEEEeqnarray}{rCl}
\IEEEeqnarraymulticol{3}{l}{
\Ex{}{e_m^C e_n^R} 
} \nonumber\\
&=& \frac{\Delta^2}{2\pi^2}\sum_{a = 1}^\infty \sum_{b=1}^\infty	 \frac{1}{a b} \exp\lefto( -\frac{\pi^2\lefto( a^2\sigma_m^2 + b^2\sigma_{n}^2\right)}{\Delta^2}\right) \nonumber \\
&&\times \lefto( \exp\lefto(\frac{2\pi^2 ab \, \sigma_{m,n}^C}{\Delta^2}\right) -  \exp\lefto(-\frac{2\pi^2 ab \, \sigma_{m,n}^C}{\Delta^2}\right) \right) \IEEEeqnarraynumspace \label{eq:error2_CR_midrise} \\
&=& \frac{\Delta^2}{\pi^2}\sum_{a = 1}^\infty \sum_{b=1}^\infty	 \frac{1}{a b} \exp\lefto( -\frac{\pi^2\lefto( a^2\sigma_m^2 + b^2\sigma_{n}^2\right)}{\Delta^2}\right)  \nonumber\\
&& \times \sinh\lefto(\frac{2\pi^2 ab \, \sigma_{mn}^C}{\Delta^2}\right).\label{eq:error3_CR_midrise}
\end{IEEEeqnarray}
Here, \eqref{eq:error2_CR_midrise} holds as $\Re\lefto\{ e^{j\pi\lefto( a - b\right)}\right\} = \cos(\pi(a-b)) = (-1)^{a+b}$ and $\Re\lefto\{ e^{j\pi\lefto( a + b\right)}\right\} = \cos(\pi(a+b)) = (-1)^{a+b}$ for $a \in \opZ$ and $b \in \opZ$. We obtain~\eqref{eq:error3_CR_midrise} by noting that $2\sinh(x) = e^x - e^{-x}$. From~~\eqref{eq:error3_CR_midrise} we finally obtain the desired result~\eqref{eq:Cqq_nondiag_approx_even} by noting that $\Ex{}{e_m e_n^*} = 2\big(\Ex{}{e_m^R e_n^R} + j\Ex{}{e_m^I e_n^R}\big)$, which holds as the input to the DACs is a circularly-symmetric Gaussian random~variable. 

\section{Proof of \fref{thm:conv}}
\label{app:proof_conv}

Recall that, for the case $L=2$, $\alpha = \sqrt{2P/(\Delta^2 \osr B)}$ and $\text{diag}(\vecg)$ is given by~\eqref{eq:gainmatrix_1bit}. Hence, since $P<\infty$, it follows from~\eqref{eq:uncorrcovariance_final}~that
\begin{IEEEeqnarray}{rCl} \label{eq:jag_ar_trott_nu}
	\lim_{\Delta \rightarrow \infty} \matC_\vecx &=& \lim_{\Delta \rightarrow \infty} {\frac{2P}{\Delta^2\osr B}}\matC_\vece.
\end{IEEEeqnarray}
Hence, to prove~\fref{thm:conv}, we need to show that the entry on the $m$th row and on the $n$th column of the RHS of~\eqref{eq:jag_ar_trott_nu} equals the RHS of~\eqref{eq:arcsine_siso} if we use \eqref{eq:Cqq_rounding} to evaluate $\matC_\vece$. 
Define~$u_a = a/\Delta$ and~$v_b = b/\Delta$. Then, we can write~the entry on the $m$th row and $n$th column of the RHS of~\eqref{eq:jag_ar_trott_nu}~as~follows:
\begin{IEEEeqnarray}{rCl}
\IEEEeqnarraymulticol{3}{l}{
\lim_{\ \Delta \rightarrow \infty} [\matC_\vecx]_{m,n}
} \nonumber\\
&=& \lim_{\Delta \rightarrow \infty} \frac{4P}{\pi^2 \Delta^2 \osr B} \nonumber\\
&& \times  \sum_{a=1}^\infty \sum_{b = 1}^\infty \frac{1}{u_a v_b} \exp\lefto( -\pi^2\lefto( u_a^2\sigma_m^2 + v_b^2\sigma_{n}^2\right)\right) \nonumber\\
&& \times\lefto(\sinh\lefto( 2\pi^2 u_a v_b \, \sigma_{m,n}^R\right) + j \sinh\lefto( 2\pi^2 u_a v_b \,\sigma_{m,n}^I\right)\right) \label{eq:unsolved_sum} \\
&=&  \frac{4P}{\pi^2 \osr B} \int_0^\infty\int_0^\infty  \frac{1}{u v}  \exp\lefto( -\pi^2\lefto( u^2\sigma_m^2 + v^2\sigma_{n}^2\right)\right)\IEEEeqnarraynumspace \nonumber\\
&&\times\lefto(\sinh\lefto( 2\pi^2 u v  \sigma_{m,n}^R\right) + j \sinh\lefto( 2\pi^2 u v \sigma_{m,n}^I\right)\right)  \text{d}u \, \text{d}v  \ \IEEEeqnarraynumspace \label{eq:unsolved_integral} \\
&=& \frac{2 P }{\pi \osr B} \lefto(\arcsin\bigg(\frac{\sigma_{m,n}^R}{\sigma_m\sigma_n}\bigg)  + j \arcsin\bigg(\frac{\sigma_{m,n}^I}{\sigma_m\sigma_n}\bigg)\right). \IEEEeqnarraynumspace  \label{eq:solved_integral}
\end{IEEEeqnarray}
Here, to obtain \eqref{eq:unsolved_sum}, we replaced $[\matC_\vece]_{m,n}$ with~\eqref{eq:Cqq_nondiag_approx_even}.
To obtain~\eqref{eq:unsolved_integral}, we used that \eqref{eq:unsolved_sum} is a two-dimensional Riemann sum, which, by definition, can be written as the two-dimensional integral in~\eqref{eq:unsolved_integral}. Finally, to obtain~\eqref{eq:solved_integral}, we used that
\begin{IEEEeqnarray}{rCl} 
&&\frac{2}{\pi}\int_0^\infty\int_0^\infty \frac{1}{u v} \exp\lefto( -\pi^2\lefto( u^2\sigma_m^2 + v^2\sigma_n^2 \right) \right) \nonumber\\
&&\times \sinh\lefto( 2\pi^2 u v \, \sigma_{m,n}^C\right) \text{d}u \, \text{d}v
= \arcsin\bigg( \frac{\sigma_{m,n}^C}{\sigma_m\sigma_n}\bigg). \IEEEeqnarraynumspace \label{eq:crazy_intergral}
\end{IEEEeqnarray}
We note that~\eqref{eq:solved_integral} is equal to the RHS of~\eqref{eq:arcsine_siso}, which concludes the proof.

\bibliographystyle{IEEEtran}
\begin{spacing}{.873}
\bibliography{IEEEabrv,confs-jrnls,publishers,svenbib}
\end{spacing}

\end{document}

%% file: vmr-symbols-vecbold.tex
\usepackage{amssymb}
\usepackage{amsfonts}
\usepackage{mathrsfs}
\usepackage{xspace}
\usepackage{bm}
\usepackage{upgreek}

\newcommand{\safemath}[2]{\newcommand{#1}{\ensuremath{#2}\xspace}}

\safemath{\bma}{\mathbf{a}}
\safemath{\bmb}{\mathbf{b}}
\safemath{\bmc}{\mathbf{c}}
\safemath{\bmd}{\mathbf{d}}
\safemath{\bme}{\mathbf{e}}
\safemath{\bmf}{\mathbf{f}}
\safemath{\bmg}{\mathbf{g}}
\safemath{\bmh}{\mathbf{h}}
\safemath{\bmi}{\mathbf{i}}
\safemath{\bmj}{\mathbf{j}}
\safemath{\bmk}{\mathbf{k}}
\safemath{\bml}{\mathbf{l}}
\safemath{\bmm}{\mathbf{m}}
\safemath{\bmn}{\mathbf{n}}
\safemath{\bmo}{\mathbf{o}}
\safemath{\bmp}{\mathbf{p}}
\safemath{\bmq}{\mathbf{q}}
\safemath{\bmr}{\mathbf{r}}
\safemath{\bms}{\mathbf{s}}
\safemath{\bmt}{\mathbf{t}}
\safemath{\bmu}{\mathbf{u}}
\safemath{\bmv}{\mathbf{v}}
\safemath{\bmw}{\mathbf{w}}
\safemath{\bmx}{\mathbf{x}}
\safemath{\bmy}{\mathbf{y}}
\safemath{\bmz}{\mathbf{z}}
\safemath{\bmzero}{\mathbf{0}}
\safemath{\bmone}{\mathbf{1}}

\bmdefine{\biad}{a}
\bmdefine{\bibd}{b}
\bmdefine{\bicd}{c}
\bmdefine{\bidd}{d}
\bmdefine{\bied}{e}
\bmdefine{\bifd}{f}
\bmdefine{\bigd}{g}
\bmdefine{\bihd}{h}
\bmdefine{\biid}{i}
\bmdefine{\bijd}{j}
\bmdefine{\bikd}{k}
\bmdefine{\bild}{l}
\bmdefine{\bimd}{m}
\bmdefine{\bind}{n}
\bmdefine{\biod}{o}
\bmdefine{\bipd}{p}
\bmdefine{\biqd}{q}
\bmdefine{\bird}{r}
\bmdefine{\bisd}{s}
\bmdefine{\bitd}{t}
\bmdefine{\biud}{u}
\bmdefine{\bivd}{v}
\bmdefine{\biwd}{w}
\bmdefine{\bixd}{x}
\bmdefine{\biyd}{y}
\bmdefine{\bizd}{z}

\bmdefine{\bixid}{\xi}
\bmdefine{\bilambdad}{\lambda}
\bmdefine{\bimud}{\mu}
\bmdefine{\bithetad}{\theta}
\bmdefine{\biphid}{\phi}
\bmdefine{\bideltad}{\delta}

\safemath{\bmia}{\biad}
\safemath{\bmib}{\bibd}
\safemath{\bmic}{\bicd}
\safemath{\bmid}{\bidd}
\safemath{\bmie}{\bied}
\safemath{\bmif}{\bifd}
\safemath{\bmig}{\bigd}
\safemath{\bmih}{\bihd}
\safemath{\bmii}{\biid}
\safemath{\bmij}{\bijd}
\safemath{\bmik}{\bikd}
\safemath{\bmil}{\bild}
\safemath{\bmim}{\bimd}
\safemath{\bmin}{\bind}
\safemath{\bmio}{\biod}
\safemath{\bmip}{\bipd}
\safemath{\bmiq}{\biqd}
\safemath{\bmir}{\bird}
\safemath{\bmis}{\bisd}
\safemath{\bmit}{\bitd}
\safemath{\bmiu}{\biud}
\safemath{\bmiv}{\bivd}
\safemath{\bmiw}{\biwd}
\safemath{\bmix}{\bixd}
\safemath{\bmiy}{\biyd}
\safemath{\bmiz}{\bizd}

\safemath{\bmxi}{\bixid}
\safemath{\bmlambda}{\bilambdad}
\safemath{\bmmu}{\bimud}
\safemath{\bmtheta}{\bithetad}
\safemath{\bmphi}{\biphid}
\safemath{\bmdelta}{\bideltad}

\safemath{\bA}{\mathbf{A}}
\safemath{\bB}{\mathbf{B}}
\safemath{\bC}{\mathbf{C}}
\safemath{\bD}{\mathbf{D}}
\safemath{\bE}{\mathbf{E}}
\safemath{\bF}{\mathbf{F}}
\safemath{\bG}{\mathbf{G}}
\safemath{\bH}{\mathbf{H}}
\safemath{\bI}{\mathbf{I}}
\safemath{\bJ}{\mathbf{J}}
\safemath{\bK}{\mathbf{K}}
\safemath{\bL}{\mathbf{L}}
\safemath{\bM}{\mathbf{M}}
\safemath{\bN}{\mathbf{N}}
\safemath{\bO}{\mathbf{O}}
\safemath{\bP}{\mathbf{P}}
\safemath{\bQ}{\mathbf{Q}}
\safemath{\bR}{\mathbf{R}}
\safemath{\bS}{\mathbf{S}}
\safemath{\bT}{\mathbf{T}}
\safemath{\bU}{\mathbf{U}}
\safemath{\bV}{\mathbf{V}}
\safemath{\bW}{\mathbf{W}}
\safemath{\bX}{\mathbf{X}}
\safemath{\bY}{\mathbf{Y}}
\safemath{\bZ}{\mathbf{Z}}

\safemath{\bZero}{\mathbf{0}}
\safemath{\bOne}{\mathbf{1}}
\safemath{\bDelta}{\mathbf{\Delta}}
\safemath{\bLambda}{\mathbf{\UpLambda}}
\safemath{\bPhi}{\mathbf{\Upphi}}
\safemath{\bSigma}{\mathbf{\Upsigma}}
\safemath{\bOmega}{\mathbf{\Upomega}}
\safemath{\bTheta}{\mathbf{\Uptheta}}

\bmdefine{\biAd}{A}
\bmdefine{\biBd}{B}
\bmdefine{\biCd}{C}
\bmdefine{\biDd}{D}
\bmdefine{\biEd}{E}
\bmdefine{\biFd}{F}
\bmdefine{\biGd}{G}
\bmdefine{\biHd}{H}
\bmdefine{\biId}{I}
\bmdefine{\biJd}{J}
\bmdefine{\biKd}{K}
\bmdefine{\biLd}{L}
\bmdefine{\biMd}{M}
\bmdefine{\biOd}{N}
\bmdefine{\biPd}{O}
\bmdefine{\biQd}{P}
\bmdefine{\biRd}{R}
\bmdefine{\biSd}{S}
\bmdefine{\biTd}{T}
\bmdefine{\biUd}{U}
\bmdefine{\biVd}{V}
\bmdefine{\biWd}{W}
\bmdefine{\biXd}{X}
\bmdefine{\biYd}{Y}
\bmdefine{\biZd}{Z}

\bmdefine{\biDelta}{\Delta}
\bmdefine{\biLambda}{\Lambda}
\bmdefine{\biPhi}{\Phi}
\bmdefine{\biSigma}{\Sigma}
\bmdefine{\biOmega}{\Omega}
\bmdefine{\biTheta}{\Theta}

\safemath{\bimA}{\biAd}
\safemath{\bimB}{\biBd}
\safemath{\bimC}{\biCd}
\safemath{\bimD}{\biDd}
\safemath{\bimE}{\biEd}
\safemath{\bimF}{\biFd}
\safemath{\bimG}{\biGd}
\safemath{\bimH}{\biHd}
\safemath{\bimI}{\biId}
\safemath{\bimJ}{\biJd}
\safemath{\bimK}{\biKd}
\safemath{\bimL}{\biLd}
\safemath{\bimM}{\biMd}
\safemath{\bimN}{\biNd}
\safemath{\bimO}{\biOd}
\safemath{\bimP}{\biPd}
\safemath{\bimQ}{\biQd}
\safemath{\bimR}{\biRd}
\safemath{\bimS}{\biSd}
\safemath{\bimT}{\biTd}
\safemath{\bimU}{\biUd}
\safemath{\bimV}{\biVd}
\safemath{\bimW}{\biWd}
\safemath{\bimX}{\biXd}
\safemath{\bimY}{\biYd}
\safemath{\bimZ}{\biZd}

\safemath{\bimDelta}{\biDelta}
\safemath{\bimLambda}{\biLambda}
\safemath{\bimPhi}{\biPhi}
\safemath{\bimSigma}{\biSigma}
\safemath{\bimOmega}{\biOmega}
\safemath{\bimTheta}{\biTheta}

\safemath{\setA}{\mathcal{A}}
\safemath{\setB}{\mathcal{B}}
\safemath{\setC}{\mathcal{C}}
\safemath{\setD}{\mathcal{D}}
\safemath{\setE}{\mathcal{E}}
\safemath{\setF}{\mathcal{F}}
\safemath{\setG}{\mathcal{G}}
\safemath{\setH}{\mathcal{H}}
\safemath{\setI}{\mathcal{I}}
\safemath{\setJ}{\mathcal{J}}
\safemath{\setK}{\mathcal{K}}
\safemath{\setL}{\mathcal{L}}
\safemath{\setM}{\mathcal{M}}
\safemath{\setN}{\mathcal{N}}
\safemath{\setO}{\mathcal{O}}
\safemath{\setP}{\mathcal{P}}
\safemath{\setQ}{\mathcal{Q}}
\safemath{\setR}{\mathcal{R}}
\safemath{\setS}{\mathcal{S}}
\safemath{\setT}{\mathcal{T}}
\safemath{\setU}{\mathcal{U}}
\safemath{\setV}{\mathcal{V}}
\safemath{\setW}{\mathcal{W}}
\safemath{\setX}{\mathcal{X}}
\safemath{\setY}{\mathcal{Y}}
\safemath{\setZ}{\mathcal{Z}}
\safemath{\emptySet}{\varnothing}

\safemath{\colA}{\mathscr{A}}
\safemath{\colB}{\mathscr{B}}
\safemath{\colC}{\mathscr{C}}
\safemath{\colD}{\mathscr{D}}
\safemath{\colE}{\mathscr{E}}
\safemath{\colF}{\mathscr{F}}
\safemath{\colG}{\mathscr{G}}
\safemath{\colH}{\mathscr{H}}
\safemath{\colI}{\mathscr{I}}
\safemath{\colJ}{\mathscr{J}}
\safemath{\colK}{\mathscr{K}}
\safemath{\colL}{\mathscr{L}}
\safemath{\colM}{\mathscr{M}}
\safemath{\colN}{\mathscr{N}}
\safemath{\colO}{\mathscr{O}}
\safemath{\colP}{\mathscr{P}}
\safemath{\colQ}{\mathscr{Q}}
\safemath{\colR}{\mathscr{R}}
\safemath{\colS}{\mathscr{S}}
\safemath{\colT}{\mathscr{T}}
\safemath{\colU}{\mathscr{U}}
\safemath{\colV}{\mathscr{V}}
\safemath{\colW}{\mathscr{W}}
\safemath{\colX}{\mathscr{X}}
\safemath{\colY}{\mathscr{Y}}
\safemath{\colZ}{\mathscr{Z}}

\safemath{\opA}{\mathbb{A}}
\safemath{\opB}{\mathbb{B}}
\safemath{\opC}{\mathbb{C}}
\safemath{\opD}{\mathbb{D}}
\safemath{\opE}{\mathbb{E}}
\safemath{\opF}{\mathbb{F}}
\safemath{\opG}{\mathbb{G}}
\safemath{\opH}{\mathbb{H}}
\safemath{\opI}{\mathbb{I}}
\safemath{\opJ}{\mathbb{J}}
\safemath{\opK}{\mathbb{K}}
\safemath{\opL}{\mathbb{L}}
\safemath{\opM}{\mathbb{M}}
\safemath{\opN}{\mathbb{N}}
\safemath{\opO}{\mathbb{O}}
\safemath{\opP}{\mathbb{P}}
\safemath{\opQ}{\mathbb{Q}}
\safemath{\opR}{\mathbb{R}}
\safemath{\opS}{\mathbb{S}}
\safemath{\opT}{\mathbb{T}}
\safemath{\opU}{\mathbb{U}}
\safemath{\opV}{\mathbb{V}}
\safemath{\opW}{\mathbb{W}}
\safemath{\opX}{\mathbb{X}}
\safemath{\opY}{\mathbb{Y}}
\safemath{\opZ}{\mathbb{Z}}
\safemath{\opZero}{\mathbb{O}}
\safemath{\identityop}{\opI}

\safemath{\veca}{\bma}
\safemath{\vecb}{\bmb}
\safemath{\vecc}{\bmc}
\safemath{\vecd}{\bmd}
\safemath{\vece}{\bme}
\safemath{\vecf}{\bmf}
\safemath{\vecg}{\bmg}
\safemath{\vech}{\bmh}
\safemath{\veci}{\bmi}
\safemath{\vecj}{\bmj}
\safemath{\veck}{\bmk}
\safemath{\vecl}{\bml}
\safemath{\vecm}{\bmm}
\safemath{\vecn}{\bmn}
\safemath{\veco}{\bmo}
\safemath{\vecp}{\bmp}
\safemath{\vecq}{\bmq}
\safemath{\vecr}{\bmr}
\safemath{\vecs}{\bms}
\safemath{\vect}{\bmt}
\safemath{\vecu}{\bmu}
\safemath{\vecv}{\bmv}
\safemath{\vecw}{\bmw}
\safemath{\vecx}{\bmx}
\safemath{\vecy}{\bmy}
\safemath{\vecz}{\bmz}

\safemath{\veczero}{\bmzero}
\safemath{\vecone}{\bmone}
\safemath{\vecxi}{\bmxi}
\safemath{\veclambda}{\bmlambda}
\safemath{\vecmu}{\bmmu}
\safemath{\vectheta}{\bmtheta}
\safemath{\vecphi}{\bmphi}
\safemath{\vecdelta}{\bmdelta}

\safemath{\matA}{\bA}
\safemath{\matB}{\bB}
\safemath{\matC}{\bC}
\safemath{\matD}{\bD}
\safemath{\matE}{\bE}
\safemath{\matF}{\bF}
\safemath{\matG}{\bG}
\safemath{\matH}{\bH}
\safemath{\matI}{\bI}
\safemath{\matJ}{\bJ}
\safemath{\matK}{\bK}
\safemath{\matL}{\bL}
\safemath{\matM}{\bM}
\safemath{\matN}{\bN}
\safemath{\matO}{\bO}
\safemath{\matP}{\bP}
\safemath{\matQ}{\bQ}
\safemath{\matR}{\bR}
\safemath{\matS}{\bS}
\safemath{\matT}{\bT}
\safemath{\matU}{\bU}
\safemath{\matV}{\bV}
\safemath{\matW}{\bW}
\safemath{\matX}{\bX}
\safemath{\matY}{\bY}
\safemath{\matZ}{\bZ}
\safemath{\matzero}{\bmzero}

\safemath{\matDelta}{\bDelta}
\safemath{\matLambda}{\bLambda}
\safemath{\matPhi}{\bPhi}
\safemath{\matSigma}{\bSigma}
\safemath{\matOmega}{\bOmega}
\safemath{\matTheta}{\bTheta}

\safemath{\matidentity}{\matI}
\safemath{\matone}{\matO}

\safemath{\rnda}{A}
\safemath{\rndb}{B}
\safemath{\rndc}{C}
\safemath{\rndd}{D}
\safemath{\rnde}{E}
\safemath{\rndf}{F}
\safemath{\rndg}{G}
\safemath{\rndh}{H}
\safemath{\rndi}{I}
\safemath{\rndj}{J}
\safemath{\rndk}{K}
\safemath{\rndl}{L}
\safemath{\rndm}{M}
\safemath{\rndn}{N}
\safemath{\rndo}{O}
\safemath{\rndp}{P}
\safemath{\rndq}{Q}
\safemath{\rndr}{R}
\safemath{\rnds}{S}
\safemath{\rndt}{T}
\safemath{\rndu}{U}
\safemath{\rndv}{V}
\safemath{\rndw}{W}
\safemath{\rndx}{X}
\safemath{\rndy}{Y}
\safemath{\rndz}{Z}

\safemath{\rveca}{\bimA}
\safemath{\rvecb}{\bimB}
\safemath{\rvecc}{\bimC}
\safemath{\rvecd}{\bimD}
\safemath{\rvece}{\bimE}
\safemath{\rvecf}{\bimF}
\safemath{\rvecg}{\bimG}
\safemath{\rvech}{\bimH}
\safemath{\rveci}{\bimI}
\safemath{\rvecj}{\bimJ}
\safemath{\rveck}{\bimK}
\safemath{\rvecl}{\bimL}
\safemath{\rvecm}{\bimM}
\safemath{\rvecn}{\bimN}
\safemath{\rveco}{\bomO}
\safemath{\rvecp}{\bimP}
\safemath{\rvecq}{\bimQ}
\safemath{\rvecr}{\bimR}
\safemath{\rvecs}{\bimS}
\safemath{\rvect}{\bimT}
\safemath{\rvecu}{\bimU}
\safemath{\rvecv}{\bimV}
\safemath{\rvecw}{\bimW}
\safemath{\rvecx}{\bimX}
\safemath{\rvecy}{\bimY}
\safemath{\rvecz}{\bimZ}

\safemath{\rvecxi}{\bmxi}
\safemath{\rveclambda}{\bmlambda}
\safemath{\rvecmu}{\bmmu}
\safemath{\rvectheta}{\bmtheta}
\safemath{\rvecphi}{\bmphi}

\safemath{\rmatA}{\bimA}
\safemath{\rmatB}{\bimB}
\safemath{\rmatC}{\bimC}
\safemath{\rmatD}{\bimD}
\safemath{\rmatE}{\bimE}
\safemath{\rmatF}{\bimF}
\safemath{\rmatG}{\bimG}
\safemath{\rmatH}{\bimH}
\safemath{\rmatI}{\bimI}
\safemath{\rmatJ}{\bimJ}
\safemath{\rmatK}{\bimK}
\safemath{\rmatL}{\bimL}
\safemath{\rmatM}{\bimM}
\safemath{\rmatN}{\bimN}
\safemath{\rmatO}{\bimO}
\safemath{\rmatP}{\bimP}
\safemath{\rmatQ}{\bimQ}
\safemath{\rmatR}{\bimR}
\safemath{\rmatS}{\bimS}
\safemath{\rmatT}{\bimT}
\safemath{\rmatU}{\bimU}
\safemath{\rmatV}{\bimV}
\safemath{\rmatW}{\bimW}
\safemath{\rmatX}{\bimX}
\safemath{\rmatY}{\bimY}
\safemath{\rmatZ}{\bimZ}

\safemath{\rmatDelta}{\bimDelta}
\safemath{\rmatLambda}{\bimLambda}
\safemath{\rmatPhi}{\bimPhi}
\safemath{\rmatSigma}{\bimSigma}
\safemath{\rmatOmega}{\bimOmega}
\safemath{\rmatTheta}{\bimTheta}

%% file: standard-macros.tex
\usepackage{amssymb}
\usepackage{amsfonts}
\usepackage{mathrsfs}
\usepackage{xspace}
\usepackage{bm}
\usepackage{fancyref}
\usepackage{textcomp}

\usepackage{multirow}
\usepackage{stmaryrd}

\newenvironment{textbmatrix}{	\setlength{\arraycolsep}{2.5pt}%
								\big[\begin{matrix}}{\end{matrix}\big]%
								\raisebox{0.08ex}{\vphantom{M}}}

\def\be{\begin{equation}}
\def\ee{\end{equation}}
\def\een{\nonumber \end{equation}}
\def\mat{\begin{bmatrix}}
\def\emat{\end{bmatrix}}
\def\btm{\begin{textbmatrix}}
\def\etm{\end{textbmatrix}}

\def\ba#1\ea{\begin{align}#1\end{align}}
\def\bas#1\eas{\begin{align*}#1\end{align*}}
\def\bs#1\es{\begin{split}#1\end{split}} 
\def\bg#1\eg{\begin{gather}#1\end{gather}}
\def\bml#1\eml{\begin{multline}#1\end{multline}}
\def\bi#1\ei{\begin{itemize}#1\end{itemize}}

\newcommand{\subtext}[1]{\text{\fontfamily{cmr}\fontshape{n}\fontseries{m}\selectfont{}#1}}
\newcommand{\lefto}{\mathopen{}\left}

\DeclareMathOperator{\tr}{tr}				
\DeclareMathOperator{\sign}{sgn}			
\DeclareMathOperator{\Prob}{\opP}			
\DeclareMathOperator{\Exop}{\opE}			
\DeclareMathOperator{\Varop}{\mathrm{Var}}
\DeclareMathOperator{\Covop}{\mathrm{Cov}}
\DeclareMathOperator{\rot}{\nabla\times}		

\newcommand{\Ex}[2]{\ensuremath{\Exop_{#1}\lefto[#2\right]}} 	
\newcommand{\abs}[1]{\lefto\lvert#1\right\rvert}		

\newcommand{\vecnorm}[1]{\lefto\lVert#1\right\rVert}		

\safemath{\dirac}{\delta}					
\safemath{\krond}{\dirac}					

\safemath{\upto}{\uparrow}
\safemath{\downto}{\downarrow}
\safemath{\iu}{j}							
\safemath{\ev}{\lambda}						
\safemath{\hilseqspace}{l^{2}}				
\newcommand{\banachfunspace}[1]{\setL^{#1}}	
\safemath{\hilfunspace}{\banachfunspace{2}}	

\safemath{\SNR}{\textsf{SNR}} 				
\safemath{\PAR}{\textsf{PAR}} 				
\safemath{\No}{N_0}							
\safemath{\Es}{E_s}							
\safemath{\Eb}{E_b}							
\safemath{\EbNo}{\frac{\Eb}{\No}}
\safemath{\EsNo}{\frac{\Es}{\No}}

\DeclareMathOperator{\CHop}{\ensuremath{\opH}} 
\safemath{\tvir}{\rndh_{\CHop}}				
\safemath{\tvtf}{\rndl_{\CHop}}				
\safemath{\spf}{\rnds_{\CHop}}				
\safemath{\bff}{H_{\CHop}}					

\safemath{\ircf}{r_{h}}						
\safemath{\tftvcf}{r_{s}}					
\safemath{\tfcf}{r_{l}}						
\safemath{\bfcf}{r_{H}}						

\safemath{\tcorr}{c_h}						
\safemath{\scf}{c_{s}}						
\safemath{\tfcorr}{c_{l}}					
\safemath{\fcorr}{c_{H}}						

\safemath{\mi}{I}							
\safemath{\capacity}{C}						

\safemath{\normal}{\mathcal{N}}			
\safemath{\jpg}{\mathcal{CN}}			
\safemath{\mchain}{\leftrightarrow}		
\newcommand{\AIC}{\ensuremath{\text{AIC}}} 

\safemath{\dB}{\,\mathrm{dB}}
\safemath{\dBm}{\,\mathrm{dBm}}
\safemath{\Hz}{\,\mathrm{Hz}}
\safemath{\kHz}{\,\mathrm{kHz}}
\safemath{\MHz}{\,\mathrm{MHz}}
\safemath{\GHz}{\,\mathrm{GHz}}
\safemath{\s}{\,\mathrm{s}}
\safemath{\ms}{\,\mathrm{ms}}
\safemath{\mus}{\,\mathrm{\text{\textmu}s}}
\safemath{\ns}{\,\mathrm{ns}}
\safemath{\ps}{\,\mathrm{ps}}
\safemath{\meter}{\,\mathrm{m}}
\safemath{\mm}{\,\mathrm{mm}}
\safemath{\cm}{\,\mathrm{cm}}
\safemath{\m}{\,\mathrm{m}}
\safemath{\W}{\,\mathrm{W}}
\safemath{\mW}{\, \mathrm{mW}}
\safemath{\J}{\,\mathrm{J}}
\safemath{\K}{\,\mathrm{K}}
\safemath{\bit}{\,\mathrm{bit}}
\safemath{\nat}{\,\mathrm{nat}}

\safemath{\define}{\triangleq}			

\safemath{\equivalent}{\sim}
\safemath{\distas}{\sim}					
\safemath{\sdiff}{\Delta}				

\safemath{\reals}{\mathbb{R}}
\safemath{\positivereals}{\reals_{+}}
\safemath{\integers}{\mathbb{Z}}
\safemath{\posint}{\integers_{+}}
\safemath{\naturals}{\mathbb{N}}
\safemath{\posnaturals}{\naturals_{+}}
\safemath{\complexset}{\mathbb{C}}
\safemath{\rationals}{\mathbb{Q}}

\newcommand*{\fancyrefapplabelprefix}{app}		
\newcommand*{\fancyrefthmlabelprefix}{thm}		
\newcommand*{\fancyreflemlabelprefix}{lem}		
\newcommand*{\fancyrefcorlabelprefix}{cor}		
\newcommand*{\fancyrefdeflabelprefix}{def}		
\newcommand*{\fancyrefproplabelprefix}{prop}	
\newcommand*{\fancyrefobslabelprefix}{obs}		
\newcommand*{\fancyrefalglabelprefix}{alg}		
\newcommand*{\fancyrefasmlabelprefix}{asm}	    

\frefformat{vario}{\fancyrefseclabelprefix}{Section~#1}
\frefformat{vario}{\fancyrefthmlabelprefix}{Theorem~#1}
\frefformat{vario}{\fancyreflemlabelprefix}{Lemma~#1}
\frefformat{vario}{\fancyrefcorlabelprefix}{Corollary~#1}
\frefformat{vario}{\fancyrefdeflabelprefix}{Definition~#1}
\frefformat{vario}{\fancyrefobslabelprefix}{Observation~#1}
\frefformat{vario}{\fancyrefasmlabelprefix}{Assumption~#1}
\frefformat{vario}{\fancyreffiglabelprefix}{Fig.~#1}
\frefformat{vario}{\fancyrefapplabelprefix}{Appendix~#1} 
\frefformat{vario}{\fancyrefproplabelprefix}{Proposition~#1}
\frefformat{vario}{\fancyrefalglabelprefix}{Algorithm~#1}
\frefformat{vario}{\fancyrefeqlabelprefix}{(#1)}

%% file: defs.tex
\newcommand{\marginnote}[1]{\marginpar{\framebox{\framebox{#1}}}}
\newtheorem{thm}{Theorem}

\safemath{\dictab}{[\,\dicta\,\,\dictb\,]}

\safemath{\ysig}{\bmy}
\safemath{\ysighat}{\hat{\ysig}}
\safemath{\ysigdim}{M}
\safemath{\xsig}{\bmx}
\safemath{\xsigdim}{N}
\safemath{\nx}{n_x}
\safemath{\zsig}{\bmz}
\safemath{\zsigdim}{\ysigdim}
\safemath{\rsig}{\bmr}
\safemath{\Adict}{\bA}
\safemath{\Adicttilde}{\widetilde{\Adict}}
\safemath{\Adictdim}{\outputdim\times\xsigdim}
\safemath{\avec}{\bma}
\safemath{\avectilde}{\tilde{\avec}}
\safemath{\Bdict}{\bB}
\safemath{\Bdicttilde}{\widetilde{\Bdict}}
\safemath{\Cdict}{\bC}
\safemath{\cvec}{\bmc}
\safemath{\Ddict}{\bD}
\safemath{\Ddictdim}{\ysigdim\times\xsigdim}
\safemath{\dvec}{\bmd}
\safemath{\Ddicttilde}{\widetilde{\bD}}
\safemath{\Bonb}{\bB}
\safemath{\bvec}{\bmb}
\safemath{\Bonbdim}{\ysigdim\times\ysigdim}
\safemath{\noise}{\bmn}
\safemath{\noisedim}{\ysigim}
\safemath{\err}{\bme}
\safemath{\errdim}{\ysigdim}
\safemath{\errset}{\setE}
\safemath{\nerr}{n_e}
\safemath{\delop}{\bP_\errset}
\safemath{\delopc}{\bP_{{\errset}^c}}

\safemath{\cplxi}{\imath}
\safemath{\cplxj}{\jmath}

\safemath{\dict}{\matD}
\safemath{\inputdim}{N}		
\safemath{\outputdim}{M}		
\safemath{\sparsity}{S}	
\safemath{\inputdimA}{{N_a}}	
\safemath{\inputdimB}{{N_b}}	
\safemath{\elemA}{{n_a}}	
\safemath{\elemB}{{n_b}}	
\safemath{\resA}{\matR_a}	
\safemath{\resB}{\matR_b}	
\safemath{\subD}{\matS} 
\safemath{\subA}{\matS_a} 
\safemath{\subB}{\matS_b} 
\safemath{\dicta}{\matA} 	
\safemath{\dictb}{\matB} 	
\safemath{\hollowS}{H}
\safemath{\hollowA}{H_a}
\safemath{\hollowB}{H_b}
\safemath{\cross}{Z}
\safemath{\coh}{\mu_d}			
\safemath{\coha}{\mu_a}			
\safemath{\cohb}{\mu_b}			
\safemath{\mubs}{\nu}	
\safemath{\cohm}{\mu_m} 
\safemath{\dictset}{\setD}	
\safemath{\dictsetp}{\dictset(\coh,\coha,\cohb)}	
\safemath{\dictsetgen}{\dictset_\text{gen}}
\safemath{\dictsetgenp}{\dictsetgen(\coh)}
\safemath{\dictsetonb}{\dictset_\text{onb}}
\safemath{\dictsetonbp}{\dictsetonb(\coh)}

\safemath{\leftside}{U}
\safemath{\rightsideA}{R_a}
\safemath{\rightsideB}{R_b}

\safemath{\indexS}{\setI_S} 

\safemath{\na}{n_a}			
\safemath{\nb}{n_b}			
\safemath{\coeffa}{p_i}	
\safemath{\coeffb}{q_j}	
\safemath{\seta}{\setP}		
\safemath{\setb}{\setQ}     
\safemath{\setw}{\setW}	
\safemath{\setz}{\setZ}	
\safemath{\cola}{\veca}		
\safemath{\colb}{\vecb}		
\safemath{\cold}{\vecd}		
\safemath{\inputvec}{\vecx} 	
\safemath{\error}{\vece}	
\safemath{\noiseout}{\vecz} 	
\safemath{\inputvecel}{x}
\safemath{\inputveca}{\vecx_a}
\safemath{\inputvecb}{\vecx_b}
\safemath{\outputvec}{\vecy}	
\safemath{\lambdamin}{\lambda_{\mathrm{min}}}

\safemath{\elltwo}{\ell_2}
\safemath{\ellone}{\ell_1}
\safemath{\ellzero}{\ell_0}
\safemath{\ellinf}{\ell_\infty}
\safemath{\ellinftilde}{\ell_{\widetilde\infty}}
\safemath{\licard}{Z(\coh,\coha,\cohb)}
\safemath{\xsol}{\hat{x}}
\safemath{\xbord}{x_b}		
\safemath{\xstat}{x_s}		
\safemath{\xstatLone}{\tilde{x}_s}
\safemath{\order}{\mathcal{O}} 
\safemath{\scales}{\Theta} 
\safemath{\ones}{\mathbf{1}} 
\safemath{\zeroes}{\mathbf{0}} 
\safemath{\thlone}{\kappa(\coh,\cohb)} 
\safemath{\constoneA}{\delta} 
\safemath{\constoneB}{\epsilon} 
\safemath{\nlarge}{L}				   
\safemath{\sumlarge}{S_\nlarge}
\safemath{\maxlarger}{P_\nlarge}	   
\safemath{\Pzero}{\textrm{P0}}	
\safemath{\Pone}{\textrm{P1}}
\safemath{\vecfir}{\vecw}			 
\safemath{\vecsec}{\vecz}
\safemath{\elvecfir}{w}              
\safemath{\elvecsec}{z}				 
\safemath{\nlargefir}{n}
\safemath{\normout}{\gamma}
\safemath{\auxfun}{h}
\safemath{\supp}{\textrm{supp}}

\safemath{\indexa}{\ell}
\safemath{\indexb}{r}
\safemath{\indexc}{i}
\safemath{\indexd}{j}

\safemath{\project}{P}